\documentclass[aps,prb,twocolumn,floatfix,showpacs,longbibliography]{revtex4-2}
\usepackage{amsmath,amssymb,graphicx,bm,color}
\usepackage{framed}
\def\veck{\mathbf k}
\def\vecq{\mathbf q}

\def\vecQ{\mathbf Q}

\def\rmi{{\rm i}}
\def\rme{{\rm e}}

\def\be{\begin{equation}}
\def\ee{\end{equation}}

\begin{document}
\title{A microscopic theory of Anderson localization of electrons in random lattices}

\author{V\'aclav  Jani\v{s} } 

\affiliation{Institute of Physics, The Czech Academy of Sciences, Na Slovance 2, CZ-18221 Praha  8,  Czech Republic}

\email{janis@fzu.cz}

\date{\today}

\begin{abstract}
The existence of Anderson localization, characterized by vanishing diffusion due to strong disorder, has been demonstrated in numerous ways. A systematic approach based on the Anderson quantum model of the Fermi gas in random lattices that can describe both diffusive and localized regimes has not yet been fully established. We build on a recent publication \cite{Janis:2025ab} and present a microscopic theory of disordered electrons that covers both the metallic phase with extended Bloch waves and the localized phase, where a propagating particle forms a quantum bound state with the hole left behind at the origin. The general theory provides a framework for constructing controlled approximations to one- and two-particle Green functions that satisfy the necessary conservation laws and causality requirements across the full range of disorder strength. It is used explicitly to derive a local, mean-field-like approximation for the two-particle irreducible vertices, enabling quantitative analysis of the solution's dynamic properties in both metallic and localized phases, including critical behavior at the mobility edge. A new instability line for the dynamical electron-hole correlation function of the metallic phase is introduced.  
\end{abstract}
\maketitle

%\begin{widetext}

\section{Introduction}

Anderson formulated a simple model of the propagation of a quantum particle in a random lattice \cite{Anderson:1958aa}. His solution, in which diffusion is suppressed over long distances when disorder is sufficiently strong, is known as Anderson localization and sparked tremendous interest in the physics community. Early research focused on locating the mobility edge between extended and localized states of the Fermi gas in models of disordered alloys \cite{Mott:1961aa,Mott:1967aa,Mott:1970aa,Thouless:1970aa,Abou-Chacra:1973aa,Thouless:1974aa}. The vanishing of diffusion was attributed to a zero of the static electric conductivity. Based on qualitative considerations and experimental evidence, Mott suggested a scenario of a discontinuous metal-insulator transition at the mobility edge with a minimal metallic conductivity \cite{Mott:1970aa,Licciardello:1975aa}. Symmetry arguments and the apparent short-range behavior of the averaged resolvent led to a conjecture that the critical behavior of interacting matrices, nonlinear $\sigma$-model, may simulate the mobility edge behavior in dimensions $2+\epsilon$ \cite{Wegner:1976aa,Wegner:1979aa,Hikami:1981aa}. 

The analogy between the Anderson localization transition and the critical behavior of phase transitions in statistical mechanics was supported by scaling arguments in Ref.~\cite{Abrahams:1979aa}. Since then, the focus has shifted from the original Anderson quantum model of disordered non-interacting electrons to classical statistical mechanical models, allowing the use of established techniques for studying the critical behavior of interacting systems, such as scaling properties and universality classes \cite{Evers:2008aa}. The models differed in the universality class \cite{Wegner:1979aa} or the underlying lattice \cite{Abou-Chacra:1973aa}. In particular, a solution on the Bethe lattice with infinite coordination number was considered a mean-field solution of the Anderson localization transition. Unlike regular lattices, the critical behavior at the localization transition was found to be exponential \cite{Efetov:1987aa,Zirnbauer:1986aa}. The exponential critical behavior was later shown to be an artifact of the Bethe lattice \cite{Mirlin:1994aa}. Moreover, the mean-field solution on the Bethe lattice was later shown to miss the correct high-dimensional limit of the Cayley tree \cite{Tarquini:2017aa,Parisi:2020aa}. The Bethe lattice is the infinite limit of a regular random graph \cite{Bollobas:1998aa}, and the concept of Anderson localization was then extended to random graphs \cite{Garcia-Mata:2017aa,Tikhonov:2019aa,Garcia-Mata:2020aa,Garcia-Mata:2022aa}.  

Random matrix models in low spatial dimensions, $d\le 2$, were known to exhibit a localized spectrum \cite{Mirlin:2000aa}. The existence of the Anderson localization transition and the mobility edge in dimensions $d>2$ was proved rigorously \cite{Frohlich:1983aa}. Numerical simulations in $d=3$ complemented the mathematical proof \cite{Kramer:1993aa,Markos:2006aa}. The concept of Anderson localization has since broadened beyond its original formulation in charge diffusion and has been extended to light propagation \cite{Segev:2013aa,Yamilov:2023aa}, the localization of thermal phonons \cite{Ni:2021aa}, and the Anderson transition in cold atoms \cite{Orso:2017aa}.
 
 All these approaches focused on the critical behavior of Anderson localization in the insulating phase. They use either localized states or a diverging scale, namely the localization length. They are inapplicable in the metallic phase, where no evident diverging scale exists. Instead, the limit of the static conductivity to zero has been used to indicate the Anderson localization transition. To find a theory of the Anderson localization transition approached from the metallic phase, the diagrammatic expansion in the Anderson model of disordered electrons, together with Green functions, was used to derive a self-consistent theory for the diffusion constant that turns zero at the localization transition \cite{Vollhardt:1980aa,Vollhardt:1980ab, Kroha:1990aa,Kroha:1990ab,Vollhardt:1992aa}. It has been assumed to be an effective mean-field-like theory, for which, however, systematic extensions have not yet been found.  
 
Over the past six decades, we have gained significant insights into the Anderson localization transition and its associated critical behavior in the localized phase, characterized by a diverging localization length. However, not all questions about the origin of the localized phase have been answered. First, the critical scale is defined only in the localized phase. Second, no order parameter analogous to those in statistical mechanics has been identified. Third, the microscopic origin of the localized states, as described by the Anderson model, remains unexplained in existing approaches.   
 
The path to a microscopic theory of Anderson localization is to use the original model of a Fermi gas on a random lattice and solve it in a controlled manner. This is achieved by diagrammatically expanding the random potential in the averaged one- and two-particle resolvents. The well-established local mean-field theory for the electronic properties of random alloys, the coherent-potential approximation (CPA) \cite{Velicky:1968aa, Elliot:1974aa}, does not lead to Anderson localization. A controlled way to go beyond the local approximation is to take the limit where the spatial dimension tends to infinity. The first fully self-consistent perturbation theory for single-electron properties, exact in infinite spatial dimensions, was shown to be equivalent to CPA \cite{Vlaming:1992aa,Janis:1992ab}. It misses vertex corrections to the Drude conductivity and is unfit to describe Anderson localization \cite{Velicky:1969aa,Khurana:1990aa}. Vertex corrections that diminish the electric conductivity are caused by backscattering and non-local correlations missed by the CPA \cite{Janis:2001aa, Janis:2003aa}. 

The charge diffusion and its vanishing depend on how far a particle can propagate in a random environment. This information is not contained in the one-particle self-energy. It is the electron-hole correlation function, derived from the two-particle Green's function, that carries information about the origin of electron propagation. It is determined from a Bethe-Salpeter equation with a two-particle irreducible vertex. The Anderson localization transition is signaled by a bifurcation point at which a new solution for the two-particle vertex emerges. Self-consistent nonlinear equations for the two-particle vertex must be used to see the bifurcation in the two-particle vertex \cite{Janis:2001ab}. We earlier used the parquet construction, combining Bethe-Salpeter equations for nonlocal irreducible vertices in the electron-hole and electron-electron scattering channels, to introduce two-particle self-consistency \cite{Janis:2005aa, Janis:2005ab}. This approach is similar to Refs.~\cite{Vollhardt:1980aa,Vollhardt:1980ab, Vollhardt:1992aa}, which replaced the full two-particle self-consistency of the parquet equations for the irreducible vertices with a self-consistent equation for the dynamical diffusion coefficient. We identified a critical point that resembles the Anderson localization transition. Our solution, however, seemed unphysical since it did not obey the necessary Ward identity between the self-energy and the irreducible electron-hole vertex of Ref.~\cite{Vollhardt:1980ab}. This Ward identity guarantees the correct low-energy asymptotics of the diffusion pole in the density response function of the metallic phase. The dynamical vertex in the perturbation expansion is causal, so the related self-energy has the demanded analytic properties. We showed that microscopic causality is incompatible with the macroscopic Ward identity \cite{Janis:2004aa,Janis:2004ab}. The observed macroscopic quantities, the electron-hole correlation function and conductivity, must conform to the conservation laws. We showed how a conserving vertex complying with the desired Ward identity is obtained from the dynamical one of the perturbation theory in Ref.~\cite{Janis:2016aa}. This addition enabled us to establish a controlled method for relating the outcome of microscopic perturbation theory to macroscopic observables. It opened a route towards a consistent microscopic theory of Anderson localization.    

I took the first step toward a fully controllable microscopic theory of Anderson localization in a recent publication \cite{Janis:2025ab}. A local approximation to the two-particle irreducible vertex in the parquet equations led to a critical point in the metallic phase of the Anderson model of disordered electrons. The critical point was identified with the Anderson localization transition. The diverging scale at the transition point was identified as the frequency derivative of the dynamical electron-hole irreducible vertex, or the dynamical conductivity. It was found to match the divergence of the localization length on the other side of the Anderson localization transition. The most fundamental conclusion of this paper was that the localized state is a quantum bound state of the propagating electron and the hole left at the beginning of its motion.   

This paper builds on Ref.~\cite{Janis:2025ab} to present a fully controllable microscopic theory of Anderson localization based on self-consistent nonlinear equations for two-particle irreducible vertices, derived from the parquet theory that interconnects multiple scattering in the electron-hole, diffuson, and electron-electron, cooperon channels. We show that, to reach the critical point signaling a transition to the localized phase, two-particle functions must contain more information than that provided by the self-energy. The localized phase contains quantum bound states characterized by a new order parameter that induces a gap in the two-particle vertex function. Because there is no condensation of the bound states, no gap appears in the one-particle self-energy. We present details of the localized bound states, determine the threshold energy for bond breaking, and describe how the localized state converts into a metal. We also discuss the relationship between the microscopic dynamical quantities derived from perturbation theory and the macroscopic observables that obey conservation laws. We further specify the extent to which the Ward identities are valid in the metallic and localized phases. We demonstrate that linear-response theory breaks down in the localized phase and that static theories cannot fully describe Anderson localization. Finally, we discuss the relationship between the dynamical conductivity, the electron-hole correlation function, and the complex diffusion function in both metallic and localized phases.

The paper is structured as follows. The basic definitions, exact relations, and restrictions required to obtain a consistent solution of the Anderson model for disordered electrons in terms of the averaged one-particle and two-particle Green functions are summarized in Sec. II. The important pieces of the derivation of a consistent mean-field approximation that yields the Anderson localization transition, presented in earlier publications (Refs.~\cite{Janis:2001ab,Janis:2003aa,Janis:2005aa,Janis:2005ab,Janis:2016aa}), are summarized in Sec. III. The detailed dynamical properties of this solution and how to obtain measurable quantities near the mobility edge and in the localized phase are discussed in Sec. IV. The impact and consequences of the presented mean-field solution on understanding Anderson localization are discussed in Sec. V.

\section{Model and fundamental relations}

\subsection{Hamiltonian and averaged resolvent}

The microscopic theory of Anderson localization of electrons in random alloys will be formulated for the  tight-binding Hamiltonian 
\begin{eqnarray}\label{eq:AD_hamiltonian}
\widehat{H} &=&\sum_{<ij>}t_{ij}\widehat{c}_{i}^{\dagger}
\widehat{c}_{j}+\sum_iV_i \widehat{c}_{i }^{\dagger } \widehat{c}_{i}\ ,
\end{eqnarray}
The atomic potential $V_{i}$ is independently and identically distributed across sites. The two terms in the Hamiltonian do not commute, and the model is fully quantum-mechanical.

We assume the validity of the ergodic hypothesis, which states that spatial averaging is equivalent to configurational averaging. Consequently, all physical quantities in the thermodynamic limit will be configurationally averaged and translationally invariant. The averaged functions of interest, which determine all relevant physical quantities, are the one-particle and two-particle resolvents in the complex energy plane. The matrix representation of the one-particle resolvent for complex energy $z=E + i\eta$ in direct space reads
%
%\begin{subequations}
\begin{align}
  \label{eq:av_1PP}
  {G}^{(1)}_{ij}(z) &=
    \left\langle\left[z\widehat{1}-\widehat{H}\right]^{-1}_{ij}
    \right\rangle_{\rm av}\,,
\end{align}
%\end{subequations}
where the angular brackets $\langle\ldots\rangle_{\rm av}$ indicate configurational averaging. Averaging restores translational invariance, and we can represent the averaged resolvent in the Hilbert space of wave vectors. Its wave-vector representation is  
\begin{align}
 \label{eq:av_1GF}
  G_{\veck}(z) &= \frac 1N\sum_{ij}\rme^{-i(\mathbf{R}_{\mathrm i} - \mathbf{R}_{\mathrm j}){\bf k}}
    \left\langle\left[z\widehat{1}-\widehat{H}\right]^{-1}_{ij}
    \right\rangle_{\rm av}
 \nonumber \\   
     &= \frac1{z - \epsilon(\veck) - \Sigma_{\veck}(z)} \,,
\end{align}
where $\epsilon(\veck)$ is the lattice dispersion relation. The second equality is the Dyson equation, which introduces the self-energy, $\Sigma_{\veck}(z)$, that contains all information about the impact of the random potential on electron propagation. The averaged one-particle Green function determines the electronic structure and spectral properties of the Fermi gas in a random lattice. It contains no information about the extent of the configurationally dependent electron wave functions at the given energy, $z = E + i0^{+}$.

\subsection{Two-particle vertex and electron-hole correlation function}

The extent of the eigenstates of the random Hamiltonian affects the two-particle resolvent or the averaged two-particle Green function. The two-particle Green function for non-interacting particles in a random potential generally has two complex energies $z_{1}$ and $z_{2}$. 
\begin{align}
  \label{eq:av_2PP}
  G^{(2)}_{ij,kl}(z_1,z_2 &)=
    \left\langle\left[z_1\widehat{1}-\widehat{H}\right]^{-1}_{ij}
    \left[z_2\widehat{1}-\widehat{H}\right]^{-1}_{kl}
    \right\rangle_{\rm av}\,.
\end{align}
The indices $i,j,k,l$ correspond to the lattice sites with positions $\mathbf R_i,\mathbf R_j,\mathbf R_k,\mathbf R_l$ respectively. 

The behavior of the two-particle Green function with energies $z_{1} \equiv E_{+} = E + \omega/2 + i0^{+}$ and $z_{2} \equiv E_{-} = E - \omega/2 - i0^{+}$ is decisive in determining whether the eigenstates of the random Hamiltonian are localized or extended. Here $E$ is the Fermi energy and $\omega$ is the transfer energy between particle and hole. These energies characterize the electron-hole resolvent $\mathcal{G}^{RA}$, the wave-vector representation of which is  
\begin{multline}\label{eq:2P_momentum}
  \mathcal{G}^{RA}_{{\bf k}{\bf k}'}(E;\omega,{\bf q})
   =  \frac1N\sum_{ijkl} 
  \rme^{-\rmi{\bf k}\cdot{\bf R}_i}
\rme^{\rmi{\bf k}' \cdot{\bf R}_j}
  \rme^{-\rmi({\bf k}' - \bf q)\cdot{\bf R}_k}
\\ \times  
  \rme^{\rmi({\bf k} - \bf q)\cdot{\bf R}_l}
 G^{(2)}_{ij,kl}
(E + \omega/2 + i 0^+,E - \omega_/2 - i 0^+)\,,
%(E_+ + i0^{+},E_- - i0^{+})
\end{multline}
with the Fermi energy $E$.
The transfer energy $\omega$ and wave vector $\vecq$ are conserved during the scattering processes when the two-particle Green function is decomposed into elementary scattering events. The dynamical variables affected by the random potential are wave vectors $\veck$ and $\veck'$. 

The averaged two-particle Green function $\mathcal{G}^{RA}_{{\bf k}{\bf k}'}(E;\omega,{\bf q})$ contains the uncorrelated propagation of two particles, the product of the averaged retarded and advanced propagators, $G^R_{\veck}(E_{+})$ and $G^A_{\veck + \vecq}(E_{-})$, with $E_{\pm} = E \pm \omega/2$, and a disorder-induced vertex correction  
\begin{multline}\label{eq:GRA-K}
\mathcal{G}^{RA}_{{\bf k}{\bf k}'}(E;\omega,{\bf q})=
G^R_{\veck}(E_{+}) G^A_{\veck - \vecq}(E_{-})
\left[N\delta_{\veck\veck'}
\right. \\ \left.
+\ \mathcal{K}^{RA}_{{\bf k}{\bf k}'}(E;\omega,{\bf q})
 G^R_{\veck'}(E_{+}) G^A_{\veck' - \vecq}(E_{-})\right]\,.
\end{multline}
The two-particle vertex $\mathcal{K}^{RA}$ obeys an integral Bethe-Salpeter equation, an analogue of the one-particle Dyson equation,  
\begin{multline} \label{eq:BS-fundamental}
\mathcal{K}^{RA}_{\mathbf{k}\mathbf{k}'}(E;\omega,\mathbf{q})
 = L^{RA}_{\mathbf{k}\mathbf{k}'}(E;\omega,\mathbf{q})
 + \frac 1N \sum_{\mathbf{k}''} L^{RA}_{\mathbf{k}\mathbf{k}''}(E;\omega,\mathbf{q}) 
 \\ \times 
   G^{R}_{\mathbf{k}''}(E_{+}) G^{A}_{\mathbf{k}'' - \vecq}( E_{-})  \mathcal{K}^{RA}_{\mathbf{k}''\mathbf{k}'}(E;\omega,\mathbf{q})  \,, 
\end{multline}
where $L^{RA}$ is the electron-hole irreducible vertex and electron-hole self-energy. The electron-hole vertex, with all its variables, is important for the microscopic description. The problem of a particle moving in an infinite lattice with random atomic potentials is solved completely if we know the one-particle self-energy $\Sigma^{R}_{\veck}(E)$ and the electron-hole irreducible vertex $L^{RA}_{\mathbf{k}\mathbf{k}'}(E;\omega,\mathbf{q})$. 

The variables of the electron-hole vertex include microscopic, dynamical variables, such as wave vectors $\veck $ and $\veck ' $, as well as macroscopic, conserved quantities: Fermi energy $E$, transfer energy $\omega$, and transfer wave vector $\vecq $. The microscopic variables cannot be neglected in the microscopic theory of Anderson localization, but they affect global, measurable quantities, such as conductivity or diffusion, only through averaging. The fundamental function for determining the measurable transport properties is the electron-hole correlation function, defined as  
\begin{equation}\label{eq:Phi-AR}
\Phi^{RA}(E;\omega,\vecq) = \frac 1{N^{2}}\sum_{\veck\veck'} \mathcal{G}^{RA}_{{\bf k}{\bf k}'}(E;\omega,{\bf q})\,.
\end{equation}
It is not, however, directly derivable from the microscopic theory. It is obtained from the electron-hole irreducible vertex   
$L^{RA}_{\mathbf{k}\mathbf{k}'}(E;\omega,\mathbf{q})$ by using the Bethe-Salpeter equation~\eqref{eq:BS-fundamental} and Eq.~\eqref{eq:GRA-K}. 

We use the Bethe-Salpeter equation~\eqref{eq:BS-fundamental} to obtain the dominant contribution to the electron-hole correlation function in the low-energy limit $\omega,q\to 0$. We determine the eigenvectors of the symmetrized version of the integral kernel in this equation. That is, we find normalized vectors $\phi^{S}_{\veck}(E)$ that satisfy the following equations
\begin{subequations}\label{eq:LRA-eigenvector}
\begin{align}
S(E)\phi^{S}_{\veck}(E) &=  \frac1{N}\sum_{\veck'}G^{R}_{\veck}(E)L^{RA}_{\veck\veck'}(E;0,\mathbf{0})
\nonumber \\ &\qquad \times
 G^{A}_{\veck'}(E)\phi^{S}_{\veck'}(E)\,,
\\
 1 &= \frac1{N}\sum_{\veck} |\phi^{S}_{\veck}(E)|^{2}\,.
\end{align}
\end{subequations}
Generally, the eigenvalues  $S(E)\le 1$ in the metallic phase.  The low-energy asymptotics of the electron-hole vertex projected to the eigenstate $\left|\phi^{S}\right\rangle$ is 
\be\label{eq:phiS-DS}
\left\langle \phi^{S}\left|G^{R}\mathcal{K}^{RA}(E;\omega,\vecq)G^{A}\right|\phi^{S}\right\rangle \to \frac{S(E)}{\mathcal{D}^{S}(E;\omega,\vecq)}
\ee
with
%\begin{subequations}
\begin{multline}\label{eq:chi-calD}
\mathcal{D}^{S}(E;\omega,\vecq) = 1 - \frac1{N^{2}}\sum_{\veck,\veck'}\phi^{S*}_{\veck}(E)G^{R}_{\veck_{+}}(E_{+})
\\ \times
L^{RA}_{\veck\veck'}(E;\omega,\mathbf{q})G^{A}_{\veck'_{-}}(E_{-}) \phi^{S}_{\veck'}(E) \,.
 \end{multline}
We assumed that the low-energy limit is dominated and controlled by the irreducible vertex $L^{RA}_{\veck\veck'}(E;\omega,\mathbf{q})$ in the denominator $\mathcal{D}^{S}(E;\omega,\vecq)$, without changing the eigenvector $\phi^{S}_{\veck}(E)$. The eigenvector $\left|\phi\right\rangle$ with the maximal eigenvalue $S(E) = 1$ then determines the low-energy asymptotics of the electron-hole correlation function. It is 
\begin{subequations}\label{eq:PhiRA-exact}
\begin{align}
\Phi^{RA}(E;\omega,\vecq) &\to \frac{|\chi(E)|^{2}}{\mathcal{D}(E;\omega,\vecq)}\,.
\end{align} 
with
\begin{align}\label{eq:chi}
\chi(E) &= \frac 1N\sum_{\veck}\phi_{\veck}(E) G^{A}_{\veck}(E) \,.
 \end{align}
\end{subequations}
It is easy to see that the electron-hole correlation function has a pole: $\mathcal{D}(E;0,\mathbf{0})=0$. It is essential, however, to know how the pole is approached as $\omega,q\to 0$.

\subsection{Ward identities, diffusion pole, and generalized Einstein relation}

We must know the dynamical electron-hole Green function $\mathcal{G}^{RA}_{\veck\veck'}(E;\omega,\vecq)$ to determine the electron-hole correlation function and the transport properties of the model. It depends on the one-particle Green functions $G^{R}_{\veck}(E_{+})$ and $G^{A}_{\veck}(E_{-})$. The solution of the two-particle functions cannot be disconnected from that of the one-particle Green function. The one-particle and two-particle functions are connected via Ward identities. There are two for the Fermi gas in random lattices.  

Velický derived the first one within the mean-field, coherent-potential approximation \cite{Velicky:1969aa}. It holds only for the zero transfer wave vector, $q=0$, and arbitrary frequency.
\be\label{eq:WI-Ve}
\frac 1N\sum_{\veck'}   \mathcal{G}^{RA}_{{\bf k}{\bf k}'}(E;\omega,{\bf 0}) = \frac1{\omega}\left[G^{R}_{\veck}(E_{+}) - G^{A}_{\veck}(E_{-}) \right] \,.
\ee
It was derived nonperturbatively but holds only if the two-particle Hilbert space can be spanned by sums of products of Bloch waves with real wave vectors $\veck$ and $\veck'$. This means that the two-particle space contains only scattering states and no quantum bound states. This Ward identity guarantees that the wave-function norm is conserved in the Hilbert space of extended states. It prevents a transition to spatially localized states.

The other Ward identity was introduced by Vollhardt and W\"olfle \cite{Vollhardt:1980ab} who proved that  
\begin{subequations}\label{eq:WI-VW}
\begin{multline}
\Delta \Sigma_{\mathbf{k}}(E;\omega,\mathbf{q})  
\\
= \frac 1{N}\sum_{\mathbf{k}'} L^{RA}_{\mathbf{k},\mathbf{k}'}(E, \omega;\mathbf{q}) \Delta G_{\mathbf{k}'}(E;\omega,\mathbf{q}) \,,
\end{multline}
where
\begin{align}
\Delta G_{\mathbf{k}}(E;\omega,\mathbf{q})  &= G^{R}_{\mathbf{k}}(E_{+}) - G^{A}_{\mathbf{k} - \vecq}(E_{-}) \,,
\\
\Delta \Sigma_{\mathbf{k}}(E;\omega,\mathbf{q})  &= \Sigma^{R}_{\mathbf{k}}(E_{+}) - \Sigma^{A}_{\mathbf{k} - \vecq}(E_{-}) \,.
\end{align}
\end{subequations}
This equality was derived from the perturbation expansion of the self-energy in the one-particle Hilbert space. It holds for arbitrary transfer frequencies $\omega$ and momenta $\vecq$. It is unrelated to the Velický identity and guarantees macroscopic particle conservation. It holds in the metallic phase, where the perturbation expansion in powers of the random potential is applicable and convergent. 

Both Ward identities have a substantial effect on the electron-hole correlation function in the metallic phase. They restrict the low-energy asymptotics to the following form 
\begin{multline}\label{eq:PhiRA-AD}
\Phi^{RA}(E;\omega,\vecq) 
\\
\doteq \frac{2\pi n(E)}{-i\omega +  A(E;\omega)\omega^{2} + D(E;\omega,\vecq)q^{2}} \,,
\end{multline}
where 
\be
n(E) = - \frac 1{N}\sum_{\veck} \int_{-\infty}^{\infty}\frac{d\omega}{\pi}f(\omega) \frac{\partial}{\partial E}\Im G^{R}_{\veck}(E + \omega)
\ee
is the density of the electron states at temperature $T$ and energy $E$, and $f(E)$ is the Fermi function. The denominator of the electron-hole correlation function contains the diffusion pole, and it vanishes in the limit $\omega\to 0$ together with $q\to 0$, as demonstrated in the preceding subsection. The linear term in frequency does not depend on the disorder strength due to particle conservation and the Ward identity~\eqref{eq:WI-VW}.  
This low-energy expansion is then controlled by two disorder-dependent functions $A(\omega)$ and $D(\omega,\vecq)$. The latter is the generalized diffusion function, and the former time scale was introduced in Ref.~\cite{Janis:2025ab}. It violates the Ward identity Eq.~\eqref{eq:WI-Ve}, which holds, however, only when no tunneling and bound two-particle states may exist. As argued in Ref.~\cite{Janis:2025ab}, Anderson localization represents a quantum bound state between the moving electron and the hole left behind, and $A(\omega)\to \infty$ as $\omega\to 0$, indicating that the tunneling barrier becomes effectively infinite and unpenetrable to infinite distances for the diffusing particle.

The diffusion function $D(\omega,\vecq)$ plays a special role in the description of diffusion. Its homogeneous, static value, the diffusion constant $D=D(E;0,\mathbf{0})$, is related to the static electrical conductivity through the Einstein relation, $\sigma = e^{2}nD$. Introducing the inhomogeneous dynamical conductivity $\sigma(E;\omega,\vecq)$, the static Einstein relation can be generalized to a dynamical form using the electron-hole correlation function in the low-energy asymptotics $\omega,q\to 0$: 
\begin{align}\label{eq:Einstein-gen}
\sigma(E;\omega,\mathbf{q}) &= \frac{- i e^{2}\omega n(E)D(E;0,\mathbf{0}) }{-i\omega +  A(E;\omega)\omega^{2}  + D(E;\omega,\vecq)q^{2}}   
\nonumber\\ 
 &= \frac{-i e^{2}\omega}{2\pi} \Phi^{RA}(E;\omega,\vecq)D(E;0,\mathbf{0}) \,.
\end{align}
The disorder-dependent coefficients in the denominator of the electron-hole correlation function in Eq.~\eqref{eq:PhiRA-AD} are proportional to second derivatives $A(E;\omega) \propto \partial^{2}\mathcal{D}(E;\omega,\mathbf{0})/\partial\omega^{2}$ and $D(E;\omega,\vecq) \propto \nabla_{\vecq}^{2}\mathcal{D}(E;\omega,\mathbf{q})$, as follows from Eq.~\eqref{eq:PhiRA-exact}. The eigenvalue $S(E) = 1$ and the first frequency derivative $\partial\mathcal{D}(E;\omega,\mathbf{0})/\partial\omega$ evaluated at $\omega=0$ remain disorder-independent, as exacted from the Ward identity~\eqref{eq:WI-VW}.  

We use this formula to assess the low-frequency limit of the homogeneous, $q=0$, dynamical conductivity  
\be
\sigma(E;\omega) \to \frac{-ie^{2}\omega n(E)D(E)}{-i\omega + A(E)\omega^{2}} = \frac{e^{2}n(E)D(E)}{1 + iA(E)\omega}\,.
\ee 
Anderson localization is generally assumed to occur when $D\to0$ for $n>0$, i.e., when the static diffusion constant vanishes. However, this is not a critical point. The disorder-induced critical point occurs when $A\to\infty$. This means that the frequency derivative of the imaginary part of the dynamical conductivity, or the real electric polarizability, diverges. The width of the central peak of the real part of the dynamical conductivity vanishes, but not necessarily its height, the static diffusion. We argue below that this is the critical point of the Anderson localization transition. Beyond this transition, a new order parameter emerges in the denominator of the electron-hole correlation function. Consequently, the diffusion pole and diffusion vanish, as does the static conductivity $\sigma(E;0)=0$ calculated from the generalized Einstein relation, Eq.~\eqref{eq:Einstein-gen}, even if the static diffusion constant from the metallic phase is positive, $D(E)>0$. 

It is clear from the exact representation of the electron-hole correlation function, Eqs.~\eqref{eq:phiS-DS}-\eqref{eq:PhiRA-exact}, that one cannot derive a closed equation for either parameter $A(E;\omega)$ or $D(E;\omega,\vecq)$. The complete microscopic theory must give, apart from the one-electron self-energy $\Sigma^{R}_{\veck}(E)$, expressions for the full irreducible vertex $L^{RA}_{\veck\veck'}(E;\omega,\mathbf{q})$ and the eigenvector $\phi_{\veck}(E)$. The theory of Vollhardt and W\"olfle, \cite{Vollhardt:1980aa,Vollhardt:1980ab, Vollhardt:1992aa}, which leads to a self-consistent equation for $D(E;\omega,\mathbf{0})$, is only an effective approximation. Its derivation is not fully microscopically controlled, and it cannot be viewed as a consistent microscopic theory of Anderson localization as long as the corresponding electron-hole irreducible vertex $L^{RA}_{\veck\veck'}(E;\omega,\mathbf{q})$ remains unknown.

\section{Perturbation expansion}

The complete microscopic theory of charge diffusion and Anderson localization must determine the one-particle and two-particle Green functions in accordance with the macroscopic conservation laws. Moreover, it must yield two-particle functions that are not fully determined by the one-particle self-energy for Anderson localization to occur. We hence invert the construction of approximations suitable for describing Anderson localization and employ a bottom-up approach, in which we determine the self-energy from the approximated two-particle irreducible vertex. This will be used beyond the local mean-field theory. The mean-field irreducible vertex, determined from the local self-energy, will serve as input to the two-particle approach.

\subsection{Coherent potential approximation}

The coherent-potential approximation (CPA) is the best local approximation to the self-energy. It represents a dynamical mean-field theory for the disordered Fermi gas \cite{Janis:1989aa}. It is determined for a complex energy $z = E + i\eta$ with $\eta \neq 0$ from a Soven equation \cite{Velicky:1968aa}.
\begin{subequations}\label{eq:CPA_1}
\begin{align}
 G(z)&=\left\langle\left[G^{-1}(z) + \Sigma(z) -
V_i\right]^{-1}\right\rangle_{av} \,,
 \\
G(z) &= \frac 1N\sum_{\veck} G(z,\veck ) = \int_{-\infty}^{\infty}\frac{\rho(\epsilon)d\epsilon}{z - \epsilon - \Sigma(z)} \,, 
\end{align}
\end{subequations}
where $\rho(\epsilon)$ is the density of states. It is a single-site approximation with multiple scattering on a single atomic potential $V_{i}$. It does not include backscattering effects, which arise only when distant scattering centers are included. 

The coherent-potential approximation contains all information about the random lattice in the self-energy $\Sigma(z)$. All higher-order irreducible vertices are local and are determined from the self-energy. Consequently, the two-particle irreducible vertex $\lambda(z_{+},z_{-})$ is determined directly from the Ward identity~\eqref{eq:WI-VW}.
\begin{align}\label{eq:lambda-CPA}
\lambda(z_{+},z_{-}) &= \frac {\Sigma(z_+) - \Sigma(z_-)} {G(z_+) - G(z_-)} = \frac
    {\Delta\Sigma(E;\eta)} {\Delta G(E;\eta)}\,,
\end{align}
with $E= (z_{+} + z_{-})/2$ and $\eta = z_{+} - z_{-}$. The full two-particle nonlocal vertex in this mean-field approximation is 
\begin{subequations}
\begin{align}
  \mathcal{K}_{\veck\veck'}({z_{+},z_{-};\bf q}) &=   \frac{\lambda(z_+,z_-)}{1 -  \lambda(z_+,z_-)\chi(z_{+},z_{-};\vecq)} 
\end{align}
with the two-particle bubble
\begin{align} \label{eq:DRF_thermo}
  \chi(z_{+},z_{-};{\bf q}) =  \frac1N\sum_{\veck} G_{\veck_{+}}(z_{+})G_{\veck_{-}}(z_{-})  \,,
\end{align}
\end{subequations}
where $\veck_{\pm} = \veck \pm \vecq/2$. The local two-particle vertex is
\be\label{eq:gamma-CPA}
\gamma(z_{+},z_{-}) = \frac{\lambda(z_+,z_-)}{1 - \lambda(z_+,z_-)G(z_{+})G(z_{-})} \,.
\ee

% \begin{multline}\label{eq:CPA_2}
%    \lambda(z_+,z_-)=\frac 1{G(z_+)G(z_-)}\left[1- {\left\langle \frac 1{
%            1+\left(\Sigma(z_+)-V_i\right)G(z_+)}\frac 1{ 1+\left(
%              \Sigma(z_-)-V_i\right)G(z_-)}\right\rangle_{av}}^{-1}
%    \right]\\ = \frac {\Sigma(z_+) - \Sigma(z_-)} {G(z_+) - G(z_-)} = \frac
%    {\Delta\Sigma} {\Delta G} \ .
% \end{multline}
 
The electron-hole correlation function is 
\begin{subequations}\label{eq:PhiRA-CPA}
\begin{align}
\Phi^{RA}(E;\omega,\vecq) &= \frac{\chi^{RA}(E;\omega,\vecq)}{1 - \lambda^{RA}(E;\omega) \chi^{RA}(E;\omega,\vecq)} \,,
\\
\chi^{RA}(E;\omega,\vecq)&= \frac 1N\sum_{\veck}G^{R}_{\veck_{+}}(E_{+}) G^{A}_{\veck_{-}}(E_{-}) \,.
\end{align}
\end{subequations}

It is easy to find the eigenvector of the Bethe-Salpeter equation for the electron-hole vertex. It is $\phi_{\veck}(E) = G^{R}_{\veck}(E)/\sqrt{\left\langle |G^{R}|^{2}\right\rangle}$, with eigenvalue $S(E)= 1$, since $\lambda(E;0) = 1/\left\langle |G^{R}|^{2}\right\rangle$. The eigenvalue is $S(E)= 1$, and the eigenvector is $\phi_{\veck}(E) = G^{R}_{\veck}(E)/\sqrt{\left\langle |G^{R}|^{2}\right\rangle}$. The representation~\eqref{eq:PhiRA-exact} with this eigenvector gives the exact result for the electron-hole correlation function, Eq.~\eqref{eq:PhiRA-CPA}. 

%Notice that we can invert the construction of the averaged Green functions of the coherent-potential approximation. We can determine the imaginary part of the self-energy from the known irreducible vertex and Eq.~\eqref{eq:lambda-CPA}, that is
%\be
%\Im\Sigma^{R}(E) = \frac{\lambda^{RA}(E;0)}{N}\sum_{\veck}\Im G^{R}_{\veck}(E) =  \Im\Sigma^{R}(E)\lambda^{RA}(E;0)\left\langle |G^{R}|^{2}\right\rangle \,,
%\ee 
%where the angular brackets denote a normalized sum over the wave vectors. 
%The real part of this self-energy is obtained from the Kramers-Kronig relation, guaranteeing causality and proper analytic properties:
%%
%\begin{align}\label{eq:KK-real}
%\Re \Sigma^{R}(E) & =  P\int_{-\infty}^{\infty} \frac{d\omega}{\pi} \frac{\Im\Sigma^{R}(\omega)}{\omega - E} \,.
%\end{align}

\subsection{Beyond the local mean field}

The CPA is a good approximation for spectral one-particle properties. It, however, does not go beyond the semiclassical Drude form of the electric conductivity. We must change the strategy for calculating the averaged Green and correlation functions. We replace the self-energy $\Sigma^{R}_{\veck}(E)$ by the two-particle irreducible vertex $\Lambda_{\veck\veck'}(E;\omega,\vecq)$ as the fundamental quantity characterizing the selected approximation. The Dyson equation is replaced by a Bethe-Salpeter equation in the electron-hole channel, which determines the full two-particle vertex. 
\begin{multline}
 \Gamma^{RA}_{\mathbf{k}\mathbf{k}'}(E;\omega,\mathbf{q}) =
  {\Lambda}^{RA}_{\mathbf{k}\mathbf{k}'}(E;\omega,\mathbf{q}) + \frac
  1N\sum_{\mathbf{k}''}
  {\Lambda}^{RA}_{\mathbf{k}\mathbf{k}''}(E;\omega,\mathbf{q}) 
  \\ \quad \times
  {G}^{R}_{\mathbf{k}_{+}''}(E_{+}) {G}^{A}_{\mathbf{k}_{-}''}(E_{-})
  \Gamma^{RA}_{\mathbf{k}''\mathbf{k}'}(E;\omega,\mathbf{q}) \,.
\end{multline}
The electron-hole irreducible dynamical vertex ${\Lambda}^{RA}_{\mathbf{k}\mathbf{k}'}(E;\omega,\mathbf{q})$ will be determined from renormalized perturbation theory together with the one-particle propagators $G^{RA}_{\veck(E)}$. The dynamical vertex ${\Lambda}^{RA}_{\mathbf{k}\mathbf{k}'}(E;\omega,\mathbf{q})$ differs from the conserving vertex ${L}^{RA}_{\mathbf{k}\mathbf{k}'}(E;\omega,\mathbf{q})$ when $\omega \neq 0$ or $q \neq 0$. Their exact relation will be discussed later. It is important to stress that the dynamical vertex contains more information than the self-energy, although it must be related to the one-particle propagators. Only its static homogeneous part, $\omega = 0$ and $q=0$, determines the imaginary part of the self-energy, as demanded by the static limit of the Ward identity~\eqref{eq:WI-VW}
\begin{subequations}\label{eq:KK-nonlocal}
\begin{align}\label{eq:KK-imaginary-nonlocal}
\Im \Sigma^{R}_{\mathbf{k}}(E) & = \frac 1N\sum_{\mathbf{k}'}\Lambda^{RA}_{\mathbf{k}\mathbf{k}'}(E;0,\mathbf 0) \Im G^{R}_{\mathbf{k}}(E) \ .
\end{align}
The corresponding real part of the self-energy is then calculated from the Kramers-Kronig relation
\begin{align}\label{eq:KK-real-nonlocal}
\Re \Sigma^{R}_{\veck}(E) & =  P\int_{-\infty}^{\infty} \frac{d\omega}{\pi} \frac{\Im\Sigma^{R}_{\veck}(\omega)}{\omega - E} \,.
\end{align}
\end{subequations}
Equations~\eqref{eq:KK-nonlocal} uniquely determine the self-energy $\Sigma^{R}_{\mathbf{k}}(E)$ for the given vertex function $\Lambda^{RA}_{\mathbf{k}\mathbf{k}'}(E;0,\mathbf 0)$. Unlike the CPA, these equations determine the self-energy only for real energies.

\subsection{Parquet equations and two-particle self-consistency}

\begin{figure}\includegraphics[width=8.5cm]{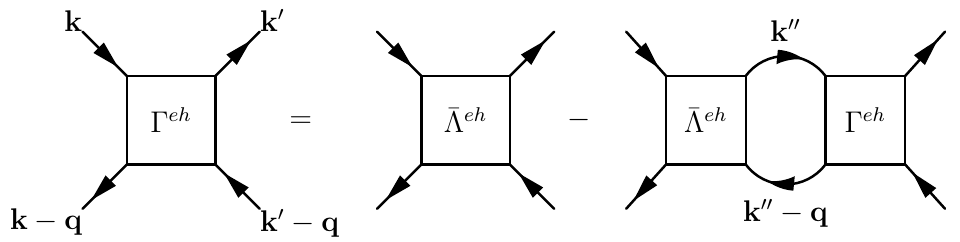}
\caption{Diagrammatic representation of the Bethe-Salpeter equation for multiple scattering in the electron-hole channel (diffuson). The two-particle vertices, labeled by wave vectors $\veck,\veck', \vecq$, are connected by off-diagonal one-electron propagators, and the double-primed wave vectors are summed over the Brillouin zone.  \label{fig:BS-eh}}
\end{figure}
\begin{figure}\includegraphics[width=8.5cm]{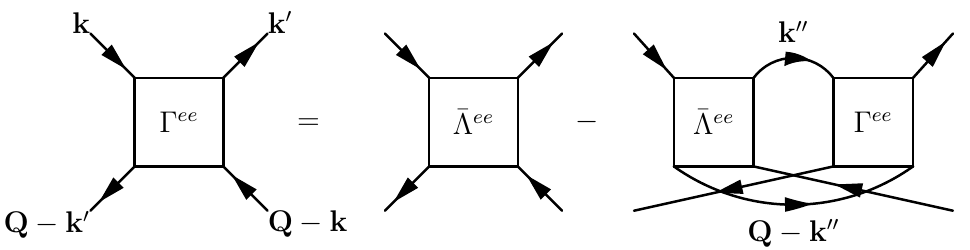}
\caption{Diagrammatic representation of the Bethe-Salpeter equation for multiple scattering in the electron-electron channel (cooperon). The two-particle vertices in this channel are characterized by wave vectors $\veck,\veck',\vecQ$. It is connected to the electron-hole channel by inverting one of the electron lines.\label{fig:BS-ee}}
\end{figure}

An important aspect of diffusion in random media that the CPA misses is backscattering. It can be included only if the approximation contains more than a single scattering center. The CPA accounts for all single-site scattering events and may serve as a suitable starting point for incorporating non-local effects and backscattering. We use the off-diagonal CPA propagator as the unperturbed Green function in the expansion beyond CPA. It is represented in the energy-wave vector space as   
\begin{equation}\label{eq:Gbar}
\overline{G}_{\mathbf{k}}( z)  = \frac{1}{z - \epsilon(\mathbf{k}) - \Sigma_{0}(z)}
 - \int_{-\infty}^{\infty}\frac{\rho(\epsilon)d\epsilon}{z - \epsilon - \Sigma(z)} \,. 
\end{equation}  
The random potential is replaced by the full local CPA vertex $\gamma(z_{+},z_{2})$ from Eq.~\eqref{eq:gamma-CPA} to determine the two-particle vertices. 

Inclusion of non-local scattering yields a crucial improvement over the local CPA. Scattering on spatially distant random potentials enables us to distinguish between the propagation of the particle, diagrammatically represented by an oriented solid line pointing from left to right, and that of the hole, represented by a solid line pointing from right to left. There is no preferred direction for the local CPA propagators. The one-particle functions can be described equivalently in the particle or the hole formalism. The situation is different at the two-particle level. The simultaneous propagation of two particles between distinct sites, shown as parallel lines in the diagrammatic representation, differs from the propagation of the particle-hole pair, shown as antiparallel lines. The two pair propagations define different two-particle irreducibility and generate two non-equivalent representations of the two-particle vertex via Bethe-Salpeter equations with non-local propagators from Eq.~\eqref{eq:Gbar}\cite{Janis:2001ab,Janis:2005aa}. The Bethe-Salpeter integral equations represent multiple scattering in the electron-hole or electron-electron channels and introduce distinct two-particle-irreducible functions. The irreducible vertex $\bar{\Lambda}^{eh}_{\veck\veck'}(E;\omega,\vecq)$ in the electron-hole channel differs from the irreducible vertex $\bar{\Lambda}^{ee}_{\veck\veck'}(E;\omega,\vecQ)$ in the electron-electron channel. We denote the bosonic vector conserved in electron-hole scattering events $\vecq$ and in the electron-electron channel $\vecQ = \veck + \veck' - \vecq$. The Bethe-Salpeter equations in the electron-hole and electron-electron channels are diagrammatically represented in Figs.~\ref{fig:BS-eh} and~\ref{fig:BS-ee}, respectively. Their mathematical form is    
\begin{subequations}\label{eq:BS}\begin{multline}
\label{eq:BS-eh}
  \Gamma^{eh}_{\mathbf{k}\mathbf{k}'}(E;\omega,\mathbf{q}) 
  \\
  =
  \bar{\Lambda}^{eh}_{\veck\veck'}(E;\omega,\mathbf{q}) 
 % \nonumber \\   &\quad 
 +\ \frac1N\sum_{\veck''}
  \bar{\Lambda}^{eh}_{\veck\veck''}(E;\omega,\mathbf{q}) %
  \\ \quad \times
  \bar{G}_{\veck''} \bar{G}_{\veck'' - \vecq} 
  \Gamma^{eh}_{\veck''\veck'}(E;\omega,\mathbf{q}) \,,
  \end{multline}
  \begin{multline}
\label{eq:BS-ee}
  \Gamma^{ee}_{\veck\veck'}(E;\omega, \vecQ) 
  \\
  =
  \bar{\Lambda}^{ee}_{\veck\veck'}(E;\omega,\vecQ) 
%  \nonumber \\ &\quad 
  + \frac 1N\sum_{\veck''}
  \bar{\Lambda}^{ee}_{\veck\veck''}(E;\omega,\vecQ) 
  \\ \quad \times
  \bar{G}_{\veck''} \bar{G}_{\vecQ - \veck''} 
  \Gamma^{ee}_{\veck''\veck'}(E;\omega,\vecQ) \,.
\end{multline}\end{subequations}
We distinguished the two-particle vertices by the Bethe-Salpeter equation that determined them and by the conserving bosonic vectors we used to characterize them. The two equations are identical, differing only in the transfer vectors. That is, $\Gamma^{eh}_{\mathbf{k}\mathbf{k}'}(E;\omega,\mathbf{q}) = \Gamma^{ee}_{\mathbf{k}\mathbf{k}'}(E;\omega,\veck + \veck'-\mathbf{q})$. 
%The Bethe-Salpeter equation in the electron-hole channel, Eq.~\eqref{eq:BS-eh},  is graphically represented in Fig.~\ref{fig:BS-eh} and in the electron-electron channel, Eq.~\eqref{eq:BS-ee}, in Fig.~\ref{fig:BS-ee}. 

The two-particle vertex is the same on the left-hand side of Eqs.~\eqref{eq:BS}, only represented in different bases. We can hence exclude it from these equations to obtain a relation between the irreducible vertices $\bar{\Lambda}^{eh}_{\veck\veck'}(E;\omega,\mathbf{q})$ and $\bar{\Lambda}^{ee}_{\veck\veck'}(E;\veck + \veck' - \vecq)$. They are not identical beyond the CPA. They stand for the sum of all {\it irreducible diagrams} in the respective channel. The second terms on the right-hand sides of Eqs.~\eqref{eq:BS} are sums of all {\it reducible diagrams} in the respective channel. The two-particle irreducibility means that the diagram cannot be disconnected into separate parts by cutting two particle lines, whereby the mutual orientation of the lines is essential.  Each diagram is either irreducible or reducible in the selected channel. Since the Bethe-Salpeter equations in the electron-hole and electron-electron channels are non-equivalent,  no diagram can be reducible in both channels.  Consequently, the reducible diagrams from one channel are simultaneously irreducible in the other \cite{Janis:2009aa}. Notice that the electrons and holes are indistinguishable in the local static approximation, and the two Bethe-Salpeter equations are identical.  If $I_{\mathbf{k}\mathbf{k}'}(E;\omega,\mathbf{q})$ is the sum of all irreducible diagrams in both channels, we obtain a parquet equation in the basis with $\vecq$ being the conserved vector in the electron-hole channel \cite{Janis:2001ab}  
\begin{multline}\label{eq:parquet-equation}
\Gamma_{\mathbf{k}\mathbf{k}'}(E;\omega,\mathbf{q}) =
\bar{\Lambda}^{eh}_{\mathbf{k}\mathbf{k}'}(E;\omega,\mathbf{q}) 
\\
+\ \bar{\Lambda}^{ee}_{\mathbf{k}\mathbf{k}'}(E;\omega,\veck + \veck'- \vecq)
 - I_{\mathbf{k}\mathbf{k}'}(E;\omega,\mathbf{q}) \,.
\end{multline}

We replace the fully irreducible veretex $I_{\mathbf{k}\mathbf{k}'}(E;\omega,\mathbf{q})$ by the local CPA vertex $\gamma(E;\omega)$ in what follows.  The Bethe-Salpeter equations~\eqref{eq:BS} together with Eq.~\eqref{eq:parquet-equation} lead to a set of two parquet equations determining the irreducible vertices $\bar{\Lambda}^{eh}_{\veck\veck'}(E;\omega,\mathbf{q})$ and $\bar{\Lambda}^{ee}_{\veck\veck'}(E;\omega, \mathbf{Q})$ from its input, the local CPA vertex $\gamma(E;\omega)$:
\begin{subequations}\label{eq:Irreducible-parquets}
\begin{multline}\label{eq:Irreducible-parquets-eh}
\frac 1N\sum_{\veck''}\left[\delta_{\veck,\veck''} - \bar{\Lambda}^{eh}_{\veck\veck''}(E;\omega,\mathbf{q})
  \bar{G}_{\veck''} \bar{G}_{\veck'' - \vecq} \right]
  \\ \times
  \bar{\Lambda}^{ee}_{\veck''\veck'}(E;\omega,\veck + \veck''-\mathbf{q})
\\
= \frac 1N\sum_{\veck''}\left[\delta_{\veck,\veck''} + \bar{\Lambda}^{eh}_{\veck\veck''}(\mathbf{q})
  \bar{G}_{\veck''}(E_{+}) \bar{G}_{\veck'' - \vecq}(E_{-}) \right]
  \\ \times
  \left[\bar{\Lambda}^{eh}_{\veck''\veck'}(E;\omega,\mathbf{q}) - \gamma(E;\omega)\right] \,,
\end{multline}
\begin{multline}\label{eq:Irreducible-parquets-ee}
\frac 1N\sum_{\veck''}\left[\delta_{\veck,\veck''} - \bar{\Lambda}^{ee}_{\veck\veck''}(E;\omega,\mathbf{Q})
  \bar{G}_{\veck''}(E_{+}) \bar{G}_{\vecQ - \veck''}(E_{-}) \right]
  \\ \times 
  \bar{\Lambda}^{eh}_{\veck''\veck'}(\veck + \veck'' - \mathbf{Q})
  \\
= \frac 1N\sum_{\veck''}\left[\delta_{\veck,\veck''} + \bar{\Lambda}^{ee}_{\veck\veck''}(E;\omega,\mathbf{Q})
  \bar{G}_{\veck''}(E_{+}) \bar{G}_{\vecQ - \veck''}(E_{-)} \right]
  \\ \times
  \left[\bar{\Lambda}^{ee}_{\veck''\veck'}(E;\omega,\mathbf{Q}) - \gamma(E;\omega)\right]
\end{multline}
\end{subequations}
%We remind that we used different representations of the two-particle irreducible vertices in these equations to characterize them with the conserving vector in the respective scattering channel. 

We next use the time reversal symmetry, that is, $G_{\vecQ -\veck''}(E_{-}) = G_{\veck''- \vecQ}(E_{-})$ in Eq.~\eqref{eq:Irreducible-parquets-ee}, which transforms it into Eq.~\eqref{eq:Irreducible-parquets-eh} with $\vecQ \to\vecq$ and  $\bar{\Lambda}^{ee}_{\veck\veck^{\prime}}(E;\omega,\vecq) = \bar{\Lambda}^{eh}_{\veck\veck^{\prime}}(E;\omega, \vecq)$.  
%It is important to note that the bosonic transfer vector $\vecq$ is evaluated in different bases. Vector $\vecq$ in the electron-hole basis is used for $\Lambda^{eh}$, while the same vector $\vecq$ used in $\Lambda^{ee}$  is evaluated in the electron-electron basis and means  $\vecq = \veck + \veck'- \vecQ$ in the electron-hole basis. 
Consequently, we obtain for the full two-particle vertex from Eqs.~\eqref{eq:BS}  $\Gamma_{\mathbf{k}\mathbf{k}'}(E;\omega,\mathbf{q}) \equiv \Gamma^{eh}_{\mathbf{k}\mathbf{k}'}(E;\omega,\mathbf{q}) = \Gamma_{\mathbf{k}\mathbf{k}'}(E;\omega,\veck + \veck' - \mathbf{q})$ reflecting the electron-hole symmetry. 

From now on, we resort to a single basis with the conserving vector $\vecq$ from the electron-hole scattering channel, see Fig.~\ref{fig:BS-eh} 
%and ${\Gamma}_{\mathbf{k}\mathbf{k}'}(E;\omega,\mathbf{q}) = {\Gamma}^{eh}_{\veck\veck^{\prime}}(E;\omega,\vecq)$. 
We then have a single irreducible vertex 
$\bar{\Lambda}_{\mathbf{k}\mathbf{k}'}(E;\omega,\mathbf{q}) = \bar{\Lambda}^{eh}_{\veck\veck^{\prime}}(E;\omega,\vecq)= \bar{\Lambda}^{ee}_{\veck\veck^{\prime}}(E;\omega,\vecq)$  obeying a non-linear integral equation 
\begin{multline}\label{eq:Irreducible-vertex}
 \frac 1N\sum_{\veck''}\left[\delta_{\veck,\veck''} - \bar{\Lambda}_{\veck\veck''}(E;\omega,\mathbf{q})
  \bar{G}_{\veck''} \bar{G}_{\veck'' - \vecq} \right]
  \\ \times 
  \bar{\Lambda}_{\veck''\veck'}(E;\omega,\veck + \veck''-\mathbf{q})
\\
= \frac 1N\sum_{\veck''}\left[\delta_{\veck,\veck''} + \bar{\Lambda}_{\veck\veck''}(\mathbf{q})
  \bar{G}_{\veck''}(E_{+}) \bar{G}_{\veck'' - \vecq}(E_{-}) \right]
  \\ \times 
  \left[\bar{\Lambda}_{\veck''\veck'}(E;\omega,\mathbf{q}) - \gamma(E;\omega)\right]
\,.
\end{multline}
The parquet equation does not allow for selecting a characteristic conserving bosonic vector, as the direct electron-hole scattering on the right-hand side is mixed with the crossed electron-electron scattering on the left-hand side.

\subsection{High-dimensional approximation}

It isn't easy to solve the parquet equations~\eqref{eq:Irreducible-parquets} or~\eqref{eq:Irreducible-vertex} without approximations. It is more important to know whether these equations can lead to a critical point and a transition to a localized state than to solve them exactly. We hence resort to an approximation justified in the asymptotic limit of infinite spatial dimension, with a control parameter $d^{-1}$. A local approximation for the irreducible vertex $\bar{\Lambda}$ will guarantee the accuracy of the approximation to leading order in $d^{-1}$.

 We replace $ \bar{\Lambda}^{eh}_{\veck\veck''}(E;\omega,\mathbf{q})$   with a constant $\bar{\Lambda}_{0}(E;\omega)$ in Eq.~\eqref{eq:Irreducible-parquets-eh}. The irreducible vertex $\bar{\Lambda}^{ee}_{\mathbf{k}\mathbf{k}'}(E;\omega,\mathbf{q})$ has then the following solution
%\begin{subequations}
\begin{multline}\label{eq:Lambdakq-bar}
\bar{\Lambda}^{ee}_{\mathbf{k}\mathbf{k}'}(E;\omega,\mathbf{q}) = \gamma(E;\omega) 
 \\ %\quad
+\ \bar{\Lambda}_{0}(E;\omega)^{2} \frac{\bar{\chi}(E;\omega,\vecq)}{1 - \bar{\Lambda}_{0}(E;\omega)\bar{\chi}(E;\omega,\vecq)}\,,
\end{multline}
with 
$\bar{\chi}(E;\omega,\vecq) = N^{-1}\sum_{\veck}G^{R}_{\veck}(E_{+}) G^{A}_{\veck - \vecq}(E_{-}) - G^{R}(E_{+}) G^{A}(E_{-}) $., with the one-particle propagators containing the self-energy calculated from Eqs.~\eqref {eq:KK-nonlocal}. 

%and $\bar{\Lambda}_{0} = \Lambda_{0}/(1 - \Lambda_{0}\left\langle \chi\right\rangle)$, where $\Lambda_{0}$ is the local irreducible vertex with full one-particle propagators.  

%\be
%\Lambda^{ee}(\vecq) = \lambda + \Lambda_{0}^{2} \frac{\bar{\chi}(\vecq)}{1 - \Lambda_{0}\chi(\vecq)}
%\ee

In the next step, we close the equation for the local vertex $\bar{\Lambda}_{0}(E;\omega)$. We use the identity $\bar{\Lambda}^{ee}_{\mathbf{k}\mathbf{k}'}(E;\omega,\mathbf{q}) = \bar{\Lambda}^{eh}_{\mathbf{k}\mathbf{k}'}(E;\omega,\mathbf{q})$. The solution of Eq.~\eqref{eq:Lambdakq-bar} yields a non-local electron-hole irreducible vertex. The self-consistent equation for the local vertex $\bar{\Lambda}_{0}(E;\omega)$ is obtained by projecting Eq.~\eqref{eq:Lambdakq-bar} onto a spatially local form.  
 %
% It does not generate a new divergence in the parguet equations, and  a mean-field solution is qualitatively correct  above the upper critical dimension  $D>d_{u} = 4$ with the frequency-dependent local irreducible vertex $\Lambda_{0}(\omega)$
%
%\begin{subequations}\label{eq:Lambda0-def}
\begin{multline}\label{eq:Lambdabar0-gen}
\bar{\Lambda}_{0}(E;\omega) = \frac 1{N^{3}}\sum_{\veck\veck',\vecq} \bar{\Lambda}^{eh}_{\mathbf{k}\mathbf{k}'}(E;\omega,\mathbf{q})  = 
\gamma(E;\omega)
%\nonumber \\ \quad
\\
 +\  \frac{\bar{\Lambda}_{0}(E;\omega)^{2}}N\sum_{\vecq}\frac{\bar{\chi}(E;\omega,\vecq)}{1 - \bar{\Lambda}_{0}(E;\omega)\bar{\chi}(E;\omega,\vecq)} \,.
\end{multline} 
%or equivalently for the irreducible vertex
%%
%\begin{align}\label{eq:Lambda0-gen}
%\Lambda_{0}(E;\omega) & = \lambda(E;\omega) + \Lambda_{0}(E;\omega)^{2}\left\langle \frac{\bar{\chi}(E;\omega,\vecq)}{1 - \Lambda_{0}(E;\omega)\chi(E;\omega,\vecq)}\right\rangle_{\vecq} \,.
%\end{align}
%\end{subequations}
%

The irreducible vertex $\Lambda_{0}(E;\omega)$, the static limit of which is needed for the self-energy, is obtained in analogy to the CPA, 
\be\label{eq:Lambda-Lambdabar}
\Lambda_{0}(E;\omega) = \frac{\bar{\Lambda}_{0}(E;\omega)}{1 + \bar{\Lambda}_{0}(E;\omega)G^{R}(E_{+})G^{A}(E_{-})}\,.
\ee
The Ward identity for this local approximation determines the imaginary part of the self-energy as
\begin{multline}\label{eq:ISigma-RLambda0}
\Im\Sigma(E_{+}) = \frac{\Re\Lambda_{0}(E;0)}N\sum_{\veck}\Im G_{\veck} 
%\nonumber \\ &
= \Re\Lambda_{0}(E;0)\Im\Sigma(E_{+})
\\ \times
\int_{-\infty}^{\infty}\frac{\rho(\epsilon)d\epsilon}{\left(E - \epsilon - \Re\Sigma(E)\right)^{2} + \Im\Sigma(E_{+})^{2}}\,.
\end{multline}
We stress that all the one-particle Green functions, even those in the CPA vertex $\gamma[G]$, contain the above self-energy. 

We can replace Eq.~\eqref {eq:Lambdabar0-gen} by an alternative 
\begin{multline}\label{eq:Lambda0-gen}
\Lambda_{0}(E;\omega)  = \lambda(E;\omega) 
%\nonumber \\ &
\\
+ \Lambda_{0}(E;\omega)^{2}\left\langle \frac{\bar{\chi}(E;\omega,\vecq)}{1 - \Lambda_{0}(E;\omega)\chi(E;\omega,\vecq)}\right\rangle_{\vecq} \,,
\end{multline}
in this fully self-consistent solution, we denote $\left\langle f(\vecq)\right\rangle_{\vecq} \equiv N^{-1}\sum_{\vecq}f(\vecq)$, a summation over the transfer wave vectors. The self-consistent solution guarantees the existence of the diffusion pole at $\omega=0$, since $\Lambda_{0}(E;0)\chi(E;\mathbf{0}) = 1$ and similarly $\bar{\Lambda}_{0}(E;0)\bar{\chi}(E;\mathbf{0}) = 1$, as a direct consequence of Eqs.~\eqref{eq:Lambda-Lambdabar} and~\eqref{eq:ISigma-RLambda0}.

We know from Ref.~\cite{Janis:2025ab} that Eq.~\eqref{eq:Lambdabar0-gen} should contain a critical point at which $\partial \bar{\Lambda}_{0}(E;\omega)/\partial \omega$ diverges at $\omega=0$. It is easy to demonstrate it
\begin{multline}\label{eq:DLambdabar}
\left[1 - \bar{\Lambda}_{0}(E)\left\langle \frac{\bar{\chi}(E;\vecq)\left(2 - \bar{\Lambda}_{0}(E)\bar{\chi}(E;\vecq)\right)}{\left(1 - \bar{\Lambda}_{0}(E)\bar{\chi}(E;\vecq)\right)^{2}}\right\rangle_{\vecq} 
\right] 
\\ \times
\left.\frac{\partial \bar{\Lambda}_{0}(E;\omega)}{\partial \omega}\right|_{\omega=0} 
= \left.\frac{\partial \gamma(E;\omega)}{\partial \omega}\right|_{\omega=0} +  \bar{\Lambda}_{0}(E)^{2}
\\ \times
\left\langle \frac{1}{\left(1 - \bar{\Lambda}_{0}(E)\bar{\chi}(E;\vecq)\right)^{2}} \left.\frac{\partial \bar{\chi}(E;\omega,\vecq)}{\partial \omega}\right|_{\omega=0}\right\rangle_{\vecq}
\end{multline}
This dynamical local approximation cannot be directly continued below the upper critical dimension $d_{u}=4$. The wave-vector integrals on both sides of Eq.~\eqref{eq:DLambdabar} diverge for $d\le 4$. However, we can keep the local irreducible vertex and the electron-hole bubble static in the denominator of the integrand on the right-hand side of Eq.~\eqref{eq:Lambdabar0-gen}. In that case, we can continue the equation
\begin{multline}\label{eq:Lambda0-static}
\bar{\Lambda}_{0}(E;\omega) = \gamma(E;\omega) 
\\
+\ \bar{\Lambda}_{0}(E;\omega)^{2}\left\langle \frac{\bar{\chi}(E;\omega,\vecq)}{1 - \bar{\Lambda}_{0}(E;0)\bar{\chi}(E;0,\vecq)}\right\rangle_{\vecq} \,,
\end{multline}
down do the lower critical dimension $d_{l}=2$. There is no metallic solution below the lower critical dimension. The behavior of the localized state in $d\le2$ will be analyzed in a separate publication.

\subsection{Mean-field critical behavior in dimensions $d>2$} 
\label{sec:MFT-d>2}

The local approximation to the electron-hole irreducible vertex is justified in high spatial dimensions, suppressing wave-vector convolutions. The small parameter in the $d\to\infty$ limit of Eqs.~\eqref{eq:Lambda0-static} is the bubble $\bar{\chi}(\vecq)$. We expand the integrand in powers of this function to simplify the self-consistent equation for the local vertex $\Lambda_{0}(E;\omega)$. In leading order, we obtain a cubic equation 
\begin{align}
\bar{\Lambda}_{0}(E;\omega) &= \gamma(E;\omega) + \bar{\Lambda}_{0}(E;\omega)^{3}\left\langle \bar{\chi}(E;\omega,\vecq)^{2}\right\rangle_{\vecq} \,.
\end{align}
or equivalently
\begin{multline}
\Lambda_{0}(E;\omega) = \lambda(E;\omega) + \frac{\Lambda_{0}(E;\omega)^{3}}{1 - \Lambda_{0}(E;\omega)G^{R}(E_{+})G^{A}(E_{-})}
\\ \times 
\left[\left\langle\chi(E;\omega,\vecq)^{2}\right\rangle_{\vecq} - \left\langle \chi(E;\omega,\vecq)\right\rangle_{\vecq}^{2}\right] \,.
\end{multline}

The qualitative, mean-field-like behavior will be guaranteed if we replace the vertex $\Lambda_{0}(E;0)$ in the denominator of the second term on the right-hand side of Eq.~\eqref{eq:Lambda0-static} with the CPA irreducible vertex $\lambda(E;0)$ and the CPA bubble ${\chi}_{0}(E;0,\vecq)$ to preserve the diffusion pole. In this simplification, we do not alter the qualitative behavior of the vertex $\Lambda_{0}(E;\omega)$ in $d > 4$. The irreducible vertex is then determined from a quadratic equation.
\begin{equation}\label{eq:parquet-high_dim}
{\Lambda}_0(E;\omega) = \lambda(E;\omega) + a(E;\omega) {\Lambda}_0(E;\omega)^2
\end{equation}
with 
\begin{subequations}
\begin{align}
a(E;\omega) &= \left\langle \frac{\bar{\chi}(E;\omega,\vecq)}{1 - \lambda(E;0)\chi_{0}(E;0,\vecq)}\right\rangle_{\vecq}
\end{align}
or in the weak-disorder limit and in dimensions $d>4$ after expanding the denominator in Eq.~\eqref{eq:Lambdabar0-gen}, we obtain
\begin{multline}
a(E;\omega) = \lambda(E;\omega) 
\\ \times 
\left[\left\langle\chi(E;\omega,\vecq)^{2}\right\rangle_{\vecq} - \left\langle \chi(E;\omega,\vecq)\right\rangle_{\vecq}^{2}\right] \,.
\end{multline}
\end{subequations}
%in dimensions $2<d\le4$, where the diffusion pole is important for the frequency derivative of vertex $\Lambda_{0}$ we choose
%\begin{align}
%a(\omega) &= 2\pi n_{F}\int_{0}^{1/l} \frac{q^{d-1}dq}{-i \omega + D q^{2}}
%\end{align}
%
%where $l$ is an appropriate high-energy cutoff and $D= \left\langle \Im G^{2} \nabla\epsilon^{2}\right\rangle/|\left\langle \Im G\right\rangle|$ is the mean-field diffusion constant.

%
Notice that $a\propto d^{-1}$ in high spatial dimensions. Its physical root in the metallic regime, $4a(E;0)\lambda(E;0) <1$, is
\begin{multline}\label{eq:Lambda0}
\Lambda_{0}(E;\omega) = \frac{1}{2a(E;\omega)}
\\ \times
\left[1 - \sqrt{1 - 4a(E;\omega)\lambda(E;\omega)}\right]\,.
\end{multline}

This solution leads to a bifurcation point at $4a(E;0)\lambda_{c}=1$, where it splits into two complex conjugate roots. It is also immediately apparent that $a(E;0) \to 0$ and $\lambda_{c}\to\infty$ as $d\to\infty$. The upper critical dimension for the vanishing of this bifurcation point is then $d=\infty$, where the CPA becomes exact.   
For the disorder strength $\lambda(E;0)>\lambda_{c}$ the irreducible vertex becomes complex at zero transfer energy $\omega\to 0$ \cite{Janis:2025ab}
\begin{multline}\label{eq:Lambda0I}
\Lambda_{0}(E;\omega\to0) 
= \frac{1}{2a(E;0)}
\\ \times 
\left[1 + i\,\mathrm{sign}(\omega) \sqrt{4a(E;0)\lambda(E;0) - 1}\right]\,.
\end{multline}
The irreducible vertex $\Lambda_{0}(E;\omega) = \Lambda_{R}(E;\omega) + i\Lambda_{I}(E;\omega)$ becomes discontinuous at $\omega=0$ with a jump in the imaginary part $\Lambda_{I}(E;\omega)$. The order parameter in the new phase is $|\Lambda_{I}(E;0)|$, and its critical exponent in the local approximation is 
\be
\left|\Lambda_{I}(E;0)\right| \propto \left(\lambda(E;0) - \lambda_{c}\right)^{1/2} \,,
\ee
for $\lambda(E;0) >\lambda_{c}$. This mean-field critical behavior holds for $d>2$ but is exact only above the upper critical dimension, $d>d_{u} =4$. Notice that $\lambda_{c}=0$ in $d\le2$.

\section{Anderson localization}

We have so far developed renormalized perturbation theory for the one-particle self-energy and the two-particle irreducible vertices within and beyond the CPA. We found a critical point at which the frequency derivative of the electron-hole irreducible vertex diverges and a new phase emerges. We have not yet proven that this new state is indeed Anderson localization. We must relate the microscopically derived new phase to the macroscopically observed properties that characterize Anderson localization. The first step in this direction is to define the conserving irreducible vertex $L_{\veck\veck'}(E;\omega,\vecq)$, which guarantees the macroscopic conservation laws and the proper low-energy asymptotics of the diffusion pole in the electron-hole correlation function.

\subsection{Dynamical vertex, conserving vertex, and the Ward identity}

A common feature of the microscopic theory of disordered and interacting electrons is that diagrammatic perturbation theory cannot guarantee the full dynamical extent of conservation laws. The dynamical Ward identity, Eq.~\eqref{eq:WI-VW}, conflicts with causality, demanding $\Lambda_{\veck\veck'}(E;0,\vecq)>0$ in the metallic phase \cite{Janis:2004ab}. Only the static form of this Ward identity, used in Eq.~\eqref{eq:KK-nonlocal}, can be guaranteed by causal perturbation theory. We then have to introduce two two-particle vertices. The dynamical electron-hole vertex from microscopic perturbation theory is determined by the Bethe-Salpeter equation, with the irreducible vertex from perturbation theory $\Lambda^{RA}_{\veck\veck'}(E;\omega,\vecq)$ and the conserving vector $\vecq$ from the electron-hole scattering channel.
\begin{multline}\label{eq:G2-momentum}
\Gamma^{RA}_{{\bf k}{\bf k}'}(E;\omega,{\bf q}) 
\\
= \Lambda^{RA}_{{\bf k}{\bf k}'}(E;\omega,{\bf q}) \left[\delta(\veck - \veck^{\prime}) %\phantom{\frac12}
 + \frac 1N \sum_{\mathbf{k}''} 
\Lambda^{RA}_{\mathbf{k}\mathbf{k}''}(E;\omega,\mathbf{q})
\right. \\ \left. \times 
G^{R}_{\veck_{+}''}(E_{+})G^{A}_{\veck''_{-}}(E_{-}) \Gamma^{RA}_{\mathbf{k}''\mathbf{k}'}(E;\omega,\mathbf{q})\right]\,.
\end{multline} 

%The critical scale at the Anderson localization transition
%\be
%Z = -\frac 1{N^{2}}\sum_{\veck,\veck'} \left. \lim_{q\to 0}\frac{\partial\Im \Lambda_{\veck\veck'}(E;\omega,\vecq)}{\partial \omega}\right|_{\omega=0} 
%\ee

%The Bethe-Salpeter equation for the conserving vertex 
%%
%\begin{equation}\label{eq:G2-momentum}
%\mathcal{K}^{RA}_{{\bf k}{\bf k}'}(E;\omega,{\bf q}) = {L}^{RA}_{{\bf k}{\bf k}'}(E;\omega,{\bf q}) \left[\delta(\veck - \veck^{\prime}) \phantom{\frac12}
%%\right. \\ \left.
% + \frac 1N \sum_{\mathbf{k}''} 
%{L}^{RA}_{\mathbf{k}\mathbf{k}''}(E;\omega,\mathbf{q}) G^{R}_{\veck''_{+}}(E_{+})G^{A}_{\veck''_{-}}(E_{-}) \mathcal{K}^{RA}_{\mathbf{k}''\mathbf{k}'}(E;\omega,\mathbf{q})\right]\,.
%\end{equation} 

The conserving electron-hole vertex $\mathcal{K}^{RA}_{{\bf k}{\bf k}'}(E;\omega,{\bf q})$ is obtained from the Bethe-Salpeter equation~\eqref{eq:BS-fundamental} using the conserving irreducible vertex $L^{RA}_{\veck\veck'}(E;\omega,\vecq)$. It is derived from the dynamical vertex $\Lambda^{RA}_{\veck\veck'}(E;\omega,\vecq)$ as follows \cite{Janis:2016aa}
\begin{multline}\label{eq:L-Lambda}
L^{RA}_{\mathbf{k}\mathbf{k}'}(E;\omega,\mathbf{q}) = \Lambda^{RA}_{\mathbf{k}\mathbf{k}'}(E;\omega,\mathbf{q}) 
 -\ \frac 1{\langle\Delta G(E;\omega,\mathbf{q})^{2}\rangle}
\\ \times 
\left[\Delta G_{\mathbf{k}}(E;\omega,\mathbf{q}) R_{\mathbf{k}'}(E;\omega,\mathbf{q}) \phantom{\frac12}
+ R_{\mathbf{k}}(E;\omega,\mathbf{q}) 
\right. \\ \times \left. 
\Delta G_{\mathbf{k}'}(E;\omega,\mathbf{q}) \phantom{\frac 12}
 - \left\langle R(E;\omega,\mathbf{q}) \Delta G(E;\omega,\mathbf{q})\right\rangle
 \right. \\ \left.\times
 \frac{\Delta G_{\mathbf{k}}(E;\omega,\mathbf{q}) \Delta G_{\mathbf{k}'}(E;\omega,\mathbf{q})}{\left\langle\Delta G(E;\omega,\mathbf{q})^{2}\right\rangle}\right] \,,
\end{multline}
where the angular brackets denote the normalized integration over the fermionic momenta $\veck$ of the enclosed functions. The difference function $\Delta G_{\mathbf{k}}(E;\omega,\mathbf{q})$ was defined in Eq.~\eqref{eq:WI-VW}. The correcting function is
\begin{multline}\label{eq:R-general}
R_{\veck}(E;\omega,\vecq) = \frac 1N\sum_{\veck'}  
\left\{
\left[ \Lambda^{RA}_{\mathbf{k}\mathbf{k}'}(E;\omega,\mathbf{q}) 
\right.\right. \\ \left.\left. 
 -\ \Lambda^{RA}_{\mathbf{k}\mathbf{k}'}(E_{+}; \mathrm{sign}(\omega)0_{+},\mathbf{0})\right]   G^{R}_{\veck'}(E_{+})
% \right. \\ \left. 
+\ \left[ \Lambda^{RA}_{\mathbf{k}\mathbf{k}'}(E;\omega,\mathbf{q}) 
\right.\right. \\ \left.\left. 
-\ \Lambda^{RA}_{\mathbf{k}\mathbf{k}'}(E_{-};\mathrm{sign}(\omega)0_{-},\mathbf{0})\right] G^{A}_{\veck'}(E_{-})
\right\} \,,
\end{multline}
where we denoted $0_{\pm} = \pm \lim_{\epsilon\to 0}|\epsilon|$. 
In the metallic phase, $R_{\veck}(E;0,\mathbf{0}) = 0$. When extending the conserving vertex into the new state beyond the bifurcation point, we keep this condition, and the two irreducible vertices are identical in the homogeneous and static limit,
\begin{subequations}
\begin{align}
\Lambda^{RA}_{\veck\veck'}(E;0,\mathbf{0}) &= {L}^{RA}_{\veck\veck'}(E;0,\mathbf{0}) \,,
\end{align}
\begin{align}
\Gamma^{RA}_{\veck\veck'}(E;0,\mathbf{0}) &= \mathcal{K}^{RA}_{\veck\veck'}(E;0,\mathbf{0})
\,.   
\end{align}
\end{subequations}
This condition is important, particularly in the localized phase, where $\Lambda^{RA}_{\veck\veck'}(E;0,\mathbf{0})$ acquires an imaginary component and $\Lambda^{RA}_{\veck\veck'}(E;\omega,\mathbf{0})$ is no longer continuous at $\omega=0$. For the same reason, the Ward identity, when continued beyond the critical point into the localized phase, must be modified to 
\begin{multline}\label{WI-VW-Ext}
\Delta \Sigma_{\veck}(E;\omega,\vecq) = \frac1{2N}\sum_{\veck'}\left[L^{RA}_{\mathbf{k}\mathbf{k}'}(E;\omega,\mathbf{q}) 
\right. \\ \left.
+ L^{RA}_{\mathbf{k}\mathbf{k}'}(E; - \omega,-\mathbf{q})\right] \Delta G_{\veck'}(E;\omega,\vecq) \,,
\end{multline}
to keep the self-energy continuous.  The perturbation expansion breaks down at the critical point as well as the derivation of the Ward identity in the form of Eq.~\eqref{eq:WI-VW}.

\subsection{Local approximation}

We now calculate the conserving vertex $\mathcal{K}^{RA}_{\veck\veck'}(E;\omega,\mathbf{q})$ for the local approximation to the dynamical irreducible vertex, ${\Lambda}_{0}(E;\omega)$, from Eq.~\eqref{eq:Lambda0}. A slightly different non-local approximation with $1/d$ corrections was used in Ref.~\cite{Janis:2025ab}, with a qualitatively similar result. The advantage of the local approximation for the irreducible electron-hole vertex is that the corresponding self-energy is local. We have
\begin{subequations}
\begin{align}
 \Lambda^{RA}_{\mathbf{k}\mathbf{k}'}(E;\omega,\mathbf{q}) &= \Lambda_{0}(E;\omega) \,,
 \\
 \Im\Sigma^{R}(E) &= \Re\Lambda_{0}(E;0) \Im G^{R}(E)  \,.
 \end{align}
 \end{subequations}
 Notice that $\Re\Lambda_{0}(E;\omega)$ is continuous at $\omega=0$. The correcting function and the conserving irreducible vertex are
 \begin{subequations}
 \begin{multline}
 R(E;\omega) 
 \\
 = \left[\Lambda_{0}(E;\omega) - \Lambda_{0}(E_{+};\mathrm{sign}(\omega)0_{+})\right]\left\langle G^{R}(E_{+})\right\rangle 
 \\
 -\ \left[\Lambda_{0}(E;\omega) - \Lambda_{0}(E_{-};\mathrm{sign}(\omega)0_{-})\right]\left\langle G^{A}(E_{-})\right\rangle \,
 \end{multline}
 and
 \begin{multline}
 L_{\veck\veck'}(E;\omega,\vecq) = \Lambda_{0}(E;\omega) - \frac{R(E;\omega)}{\left\langle\Delta G(E;\omega,\vecq)^{2}\right\rangle} 
 \\ \times
 \left[\Delta G_{\veck}(E;\omega,\vecq) + \Delta G_{\veck'}(E;\omega,\vecq)
 \right. \\ \left. 
  -\ \frac{\Delta G_{\veck}(E;\omega,\vecq)\Delta G_{\veck'}(E;\omega,\vecq)}{\left\langle\Delta G(E;\omega,\vecq)^{2}\right\rangle}\right] \,.
\end{multline}
\end{subequations}

%$R(E;0) = 0$ and $L_{\veck\veck'}(E;0,\mathbf{0}) = \Lambda_{\veck\veck'}(E;0,\mathbf{0}) = \Lambda_{0}(E;0)$

The Bethe-Salpeter equation for the full conserving two-particle has the following simple form 
\begin{multline}
\mathcal{K}_{\veck\veck'} = \Lambda_{0} - \frac{R}{\left\langle \Delta G^{2}\right\rangle}\left[\Delta G_{\veck} + \Delta G_{\veck'} - \frac{\Delta G_{\veck} \Delta G_{\veck'}}{\left\langle \Delta G^{2}\right\rangle}\right] 
\\
+\ \Lambda_{0}\left\langle G^{RA}\mathcal{K}_{\veck'}\right\rangle - \frac{R}{\left\langle \Delta G^{2}\right\rangle}\left[ \Delta G_{\veck} \left\langle  G^{RA}\mathcal{K}_{\veck'}\right\rangle \phantom{\frac12}
\right. \\ \left.
 +\ \left(1 - \frac{\Delta G_{\veck} \left\langle \Delta G\right\rangle}{\left\langle \Delta G^{2}\right\rangle}\right)\left\langle \Delta G G^{RA}\mathcal{K}_{\veck'}\right\rangle\right]
\end{multline}
where we suppressed the conserving variables $E,\omega,\vecq$ to simplify notation.  We denoted $G_{\veck}^{RA}(E;\omega,\vecq) = G_{\veck_{+}}^{R}(E_{+})G_{\veck_{-}}^{A}(E_{-})$ with $E_{\pm} = E \pm \omega/2$ and $\veck_{\pm} = \veck \pm \vecq/2$. 
We further have a matrix equation for functions $ \left\langle G^{RA}\mathcal{K}_{\veck'}\right\rangle$ and $\left\langle \Delta GG^{RA}\mathcal{K}_{\veck'}\right\rangle$
\begin{widetext}
\begin{multline}
\begin{pmatrix}
1 - \Lambda_{0}\left\langle G^{RA}\right\rangle  + \displaystyle{\frac{R}{\left\langle \Delta G^{2}\right\rangle}\left\langle \Delta GG^{RA}\right\rangle}\ ,&   \displaystyle{\frac{R}{\left\langle \Delta G^{2}\right\rangle}\left(\left\langle G^{RA}\right\rangle - \frac{\left\langle \Delta G\right\rangle}{\left\langle \Delta G^{2}\right\rangle}\left\langle \Delta G G^{RA}\right\rangle\right)}
\\
 - \Lambda_{0}\left\langle \Delta G G^{RA}\right\rangle  + \displaystyle{\frac{R}{\left\langle \Delta G^{2}\right\rangle}\left\langle \Delta G^{2}G^{RA}\right\rangle}\ ,& 1 + \displaystyle{\frac{R}{\left\langle \Delta G^{2}\right\rangle}\left(\left\langle\Delta G G^{RA}\right\rangle - \frac{\left\langle \Delta G\right\rangle}{\left\langle \Delta G^{2}\right\rangle}\left\langle \Delta G^{2} G^{RA}\right\rangle\right)}
 \end{pmatrix}
 \begin{pmatrix}
 \left\langle G^{RA}\mathcal{K}_{\veck'}\right\rangle
 \\[18pt]
 \left\langle \Delta GG^{RA}\mathcal{K}_{\veck'}\right\rangle 
 \end{pmatrix}
 \\
 =
 \begin{pmatrix}
  \Lambda_{0}\left\langle G^{RA}\right\rangle - \displaystyle{\frac{R}{\left\langle \Delta G^{2}\right\rangle}
  \left[ \Delta G_{\veck'} \left\langle G^{RA}\right\rangle + \left( 1 - \frac{\Delta G_{\veck'}\left\langle \Delta G\right\rangle}{\left\langle \Delta G^{2}\right\rangle}\right)\left\langle \Delta G G^{RA}\right\rangle
  \right]}
  \\
  \Lambda_{0}\left\langle \Delta GG^{RA}\right\rangle - \displaystyle{\frac{R}{\left\langle \Delta G^{2}\right\rangle}
  \left[ \Delta G_{\veck'} \left\langle \Delta G G^{RA}\right\rangle + \left( 1 - \frac{\Delta G_{\veck'}\left\langle \Delta G\right\rangle}{\left\langle \Delta G^{2}\right\rangle}\right)\left\langle \Delta G^{2} G^{RA}\right\rangle
  \right]} 
 \end{pmatrix} \,.
\end{multline}

The determinant of the matrix on the left-hand side is 
\begin{subequations}
\begin{multline}
\mathcal{D} = 1 - \Lambda_{0}\left\langle G^{RA}\right\rangle + \displaystyle{\frac{R}{\left\langle \Delta G^{2}\right\rangle}
  \left[2 \left\langle \Delta G G^{RA}\right\rangle  - \frac{\left\langle \Delta G\right\rangle}{\left\langle \Delta G^{2}\right\rangle}\left\langle \Delta G^{2} G^{RA}\right\rangle 
  \right]}  
  \\
  +\ \frac{\Lambda_{0}R}{\left\langle \Delta G^{2}\right\rangle}
  \left[\left\langle \Delta G G^{RA}\right\rangle\left(\left\langle  G^{RA}\right\rangle  - \frac{\left\langle \Delta G\right\rangle}{\left\langle \Delta G^{2}\right\rangle}\left\langle \Delta GG^{RA}\right\rangle\right) 
 % \right. \\ \left.
  -\ \left\langle  G^{RA}\right\rangle\left(\left\langle\Delta G  G^{RA}\right\rangle  - \frac{\left\langle \Delta G\right\rangle}{\left\langle \Delta G^{2}\right\rangle}\left\langle \Delta G^{2}G^{RA}\right\rangle\right)\right] 
\\
+\  \frac{R^{2}}{\left\langle \Delta G^{2}\right\rangle^{2}}
  \left[\left\langle \Delta G G^{RA}\right\rangle\left(\left\langle\Delta G  G^{RA}\right\rangle  - \frac{\left\langle \Delta G\right\rangle}{\left\langle \Delta G^{2}\right\rangle}\left\langle \Delta G^{2}G^{RA}\right\rangle\right) 
%  \right. \\ \left.
  -\ \left\langle\Delta G^{2}  G^{RA}\right\rangle\left(\left\langle G^{RA}\right\rangle  - \frac{\left\langle \Delta G\right\rangle}{\left\langle \Delta G^{2}\right\rangle}\left\langle \Delta G G^{RA}\right\rangle\right) 
  \right] \,, 
 \end{multline}
 that can alternatively be rewritten to 
 \begin{multline}
 \mathcal{D} = \left[1 - \Lambda_{0}
  \left\langle G^{RA} \right\rangle\right]\left[1 + \frac{R}{\left\langle \Delta G^{2}\right\rangle}\left( \left\langle\Delta G G^{RA}\right\rangle  - \frac{\left\langle \Delta G\right\rangle}{\left\langle \Delta G^{2}\right\rangle}\left\langle  \Delta G^{2} G^{RA}\right\rangle \right)\right]
\\
+  \frac{R}{\left\langle \Delta G^{2}\right\rangle}\left\langle  \Delta G G^{RA}\right\rangle\left[ 
1 + \Lambda_{0}\left(\left\langle  G^{RA}\right\rangle  - \frac{\left\langle \Delta G\right\rangle}{\left\langle \Delta G^{2}\right\rangle}\left\langle  \Delta  G G^{RA}\right\rangle\right)\right]
 \\
  +\ \frac{R^{2}}{\left\langle \Delta G^{2}\right\rangle^{2}} \left[\left\langle\Delta G G^{RA}\right\rangle\left( \left\langle\Delta G G^{RA}\right\rangle  - \frac{\left\langle \Delta G\right\rangle}{\left\langle \Delta G^{2}\right\rangle}\left\langle \Delta G^{2} G^{RA}\right\rangle \right)
 -  \left\langle\Delta G^{2} G^{RA}\right\rangle\left(\left\langle G^{RA}\right\rangle  - \frac{\left\langle \Delta G\right\rangle}{\left\langle \Delta G^{2}\right\rangle}\left\langle  \Delta G G^{RA}\right\rangle\right)
  \right] \,.
\end{multline}
\end{subequations}
\end{widetext}
We further use the following representation 
\be
G^{RA} = \frac{\Delta G}{\Delta \Sigma  + \Delta \epsilon - \omega } = \frac{\Delta G}{\Lambda_{R}\left\langle \Delta G\right\rangle + \Delta \epsilon - \omega } = \gamma\Delta G\,,
\ee
where $\Lambda_{R} = \Re\Lambda_{0}$, to simplify the determinant to 
\begin{widetext}
\begin{multline}
 \mathcal{D} = \left[1 - \Lambda_{0}\left\langle\gamma \Delta G \right\rangle + \frac{R}{\left\langle \Delta G^{2}\right\rangle} \left\langle\gamma\Delta G^{2}\right\rangle \right]\left[ 1 
  + \frac{R}{\left\langle \Delta G^{2}\right\rangle} \left(\left\langle\gamma\Delta G^{2}\right\rangle  - \frac{\left\langle \Delta G\right\rangle}{\left\langle \Delta G^{2}\right\rangle}\left\langle \gamma \Delta G^{3}\right\rangle \right)\right] 
\\
+\   \frac{R}{\left\langle \Delta G^{2}\right\rangle}\left[\left\langle\gamma\Delta G\right\rangle  - \frac{\left\langle \Delta G\right\rangle}{\left\langle \Delta G^{2}\right\rangle}\left\langle \gamma \Delta G^{2}\right\rangle \right]\left[ \Lambda_{0}\left\langle\gamma \Delta G^{2} \right\rangle - \frac{R}{\left\langle \Delta G^{2}\right\rangle} \left\langle\gamma\Delta G^{3}\right\rangle\right]\,. 
\end{multline}
 \end{widetext}
 
We can now evaluate the electron-hole correlation function, knowing the conserving vertex $\mathcal{K}_{\veck\veck'}(E;\omega,\vecq)$. Its low-energy limit, where we keep only the denominator dynamical, is  
\begin{align}\label{eq:Phi-MF}
\Phi^{RA}(E;\omega,\vecq) 
%&= \frac{\left\langle |G^{R}(E)|^{2}\right\rangle}{\mathcal{D}(E;\omega,\vecq)} 
%\nonumber \\
&= \frac{\pi n(E)}{|\Im\Sigma^{R}(E)|\mathcal{D}(E;\omega,\vecq) } \,.
\end{align}

The  dynamical conductivity from the generalized Einstein relation, Eq.~\eqref{eq:Einstein-gen}, and the electron-hole correlation function,  Eq.~\eqref{eq:Phi-MF},  is
\be\label{eq:sigmaD}
\sigma(E;\omega,\vecq) = - \frac{i e^{2} \omega\ n(E)}{2\mathcal{D}(E;\omega,\mathbf{q})}\left. \nabla_{\vecq}^{2}\mathcal{D}(E;0,\vecq)\right|_{q=0} \,.
\ee
It holds, however, only within the linear response theory.
%We further use the leading orders of the low-energy asymptotics,  $R \doteq i\Lambda_{I}\left\langle \Delta G\right\rangle\omega $ with $\Lambda_{0} = \Lambda_{R} + i \Lambda_{i}$ and $\Lambda_{R} = \Im\Sigma/\left\langle \Im G \right\rangle$. 

We will need the explicit expressions for the second wave vector derivatives 
\begin{subequations}\label{eq:Momentum derivatives}
\begin{align}
\partial_{q}^{2}\left\langle\Delta G^{2}\right\rangle &= 2 \left\langle \nabla G^{R} \nabla G^{A} \right\rangle 
\\
\partial_{q}^{2}\left\langle G^{RA}\right\rangle &= \partial_{q}^{2}\left\langle \gamma \Delta G\right\rangle =  - \left\langle
\nabla G^{R}\nabla G^{A} 
\right\rangle \,,
\\
\partial_{q}^{2}\left\langle\Delta G G^{RA}\right\rangle &=  \partial_{q}^{2}\left\langle \gamma \Delta G^{2}\right\rangle 
\nonumber \\  &
= - 2 \left\langle
\Delta G_{0}  \nabla G^{R}\nabla G^{A}
\right\rangle \,,
\\
\partial_{q}^{2}\left\langle\Delta G^{2} G^{RA}\right\rangle &= \partial_{q}^{2}\left\langle \gamma \Delta G^{3}\right\rangle 
\nonumber \\ &
= - 3 \left\langle
\Delta G_{0}^{2}  \nabla G^{R}\nabla G^{A}
\right\rangle
\nonumber \\ 
 &\qquad  -  2\left\langle \gamma \Delta G_{0}\left(\nabla\Delta G\right)^{2}\right\rangle
\end{align}
\end{subequations}
where we denoted $\nabla G = \nabla_{\veck} G_{\veck} = G_{\veck}^{2}\nabla_{\veck}\epsilon(\veck)$.

\subsubsection{Metallic phase}

We evaluate the determinant $\mathcal{D}(E;\omega,\mathbf{q})$ of the electron-hole correlation function from Eq.~\eqref{eq:Phi-MF} in the low-energy limit. We start with the homogeneous limit, $q=0$. 
\begin{multline}
\mathcal{D}(E;\omega,\mathbf{0}) = \left[ 1 - \Lambda_{0}\gamma \left\langle\Delta G \right\rangle + R\gamma  \right]
\\ \times
\left[
1 + \frac{R\gamma}{\left\langle \Delta G^{2}\right\rangle}\left(\left\langle\Delta G^{2}\right\rangle  - \frac{\left\langle \Delta G\right\rangle}{\left\langle \Delta G^{2}\right\rangle}\left\langle \Delta G^{3}\right\rangle \right)\,.
\right]
\end{multline}  
We expand the right-hand side of he above equation in frequency $\omega$
 \begin{align}
 \gamma &\doteq \frac 1{\Lambda_{R}\left\langle \Delta G \right\rangle}\left[1 + \frac{\omega}{\Lambda_{R}\left\langle \Delta G \right\rangle}\right] \,,
 \\
 \Lambda_{0}  &\doteq \Lambda_{R}  + i \dot{\Lambda}_{I}\omega \,,
 \\
 R &= i\dot{\Lambda}_{I}\left\langle \Delta G\right\rangle\omega \,.
 \end{align}
Using $\Delta G = 2i\Im G^{R}$ we obtain the leading low-frequency asymptotics up to $\omega^{2}$
\begin{subequations}
\begin{multline}
\mathcal{D}(E;\omega,\mathbf{0}) \doteq -\frac{1}{2\Lambda_{R}\left\langle \Im G\right\rangle}\left[ - i\omega 
\phantom{\frac12}\right. \\ \left.
+\ \frac{\dot{\Lambda}_{I}\omega^{2}}{\Lambda_{R}\left\langle \Im G^{2}\right\rangle^{2}}\left(\left\langle \Im G^{2}\right\rangle^{2} - \left\langle \Im G\right\rangle\left\langle \Im G^{3}\right\rangle\right) \,,
\right]
\end{multline}
that can be rewritten to
\begin{multline}
\mathcal{D}(E;\omega,\mathbf{0}) \doteq - \frac{i\omega}{2|\Im \Sigma|}\left[1 
\phantom{\frac12}\right. \\ \left.
+\ \frac{i\dot{\Lambda}_{I}\left\langle |G|^{2}\right\rangle\omega}{\left\langle |G|^{4}\right\rangle^{2}}\left(\left\langle |G|^{4}\right\rangle^{2} - \left\langle |G|^{2}\right\rangle\left\langle |G|^{6}\right\rangle\right)
\right] \,.
\end{multline}
\end{subequations}

In the next step, we evaluate the leading small wave-vector limit of the denominator $\mathcal{D}(E;\omega,\mathbf{q})$. We have, using Eqs.~\eqref{eq:Momentum derivatives}
\begin{subequations}
\begin{multline}
\nabla_{\vecq}^{2} \mathcal{D}(E;\omega,\mathbf{q}) \doteq \Lambda_{R}\left\langle \nabla G^{R}\nabla G^{A} \right\rangle
\\ \times
\left[ 1 - \frac{i\dot{\Lambda}_{I}\omega}{\Lambda_{R} \left\langle \Delta G^{2} \right\rangle}
\left( \left\langle \Delta G^{2} \right\rangle + \frac{\left\langle \Delta G\right\rangle}{\left\langle \Delta G^{2} \right\rangle}\left\langle \Delta G^{3} \right\rangle\right)\right] 
\\
-\ \frac{2i\dot{\Lambda}_{I}\omega \left\langle \Delta G \right\rangle}{\left\langle \Delta G^{2} \right\rangle} \left\langle \Delta G\nabla G^{R}\nabla G^{A} \right\rangle \,.
\end{multline}
We also use the low-frequency limit 
\begin{multline}
\nabla_{\vecq}^{2} \mathcal{D}(E;\omega,\mathbf{q}) \doteq \Lambda_{R}\left\langle \nabla G^{R}\nabla G^{A} \right\rangle
\\ \times
\left[ 1 - \frac{i\dot{\Lambda}_{I}\omega}{\Lambda_{R} \left\langle \Im G^{2} \right\rangle}
\left( \left\langle \Im G^{2} \right\rangle + \frac{\left\langle \Im G\right\rangle}{\left\langle \Im G^{2} \right\rangle}\left\langle \Im G^{3} \right\rangle\right)\right] 
\\
-\ \frac{2i\dot{\Lambda}_{I}\omega \left\langle \Im G \right\rangle}{\left\langle \Im G^{2} \right\rangle} \left\langle \Im G\nabla G^{R}\nabla G^{A} \right\rangle \,.
\end{multline}
\end{subequations}

Inserting the above representations into Eq.~\eqref{eq:sigmaD}, we obtain the low-frequency limit of the dynamical conductivity 
\begin{subequations}
\begin{multline}
\sigma(E;\omega) = \frac{e^{2}}{1 + iA(E)\omega}
\\ \times
\left[ \left\langle \Im G^{R}(E)^{2}\nabla\epsilon^{2} \right\rangle\left(1 - iA(E)\omega\right)  \phantom{\frac12}
\right. \\ \left.
-\ \frac{2i \omega\dot{\Lambda}_{I}(E;0)  \left\langle \Im G^{R}(E) \right\rangle \left\langle \Im G^{R}(E)^{3}\nabla\epsilon^{2} \right\rangle }{\Im\Sigma^{R}(E)^{2} \left\langle \Im G^{R}(E)^{2} \right\rangle}\right] \,,
\end{multline}
with 
\begin{multline}
A(E) = \frac{\dot{\Lambda}_{I}(E;0)}{\Lambda_{R}(E;0)\left\langle \Im G^{R}(E)^{2}\right\rangle^{2}}
\\ \times
\left(\left\langle \Im G^{R}(E)^{2}\right\rangle^{2} - \left\langle \Im G^{R}(E)\right\rangle\left\langle \Im G^{R}(E)^{3}\right\rangle\right) \,.
\end{multline}
The critical diverging scale at the Anderson localization transition in the local approximation, Eqs.~\eqref{eq:parquet-high_dim}-\eqref{eq:Lambda0}, is 
\begin{multline}
\dot{\Lambda}_{I}(E;0) = -\lambda(E;0)
\\
\times \left\langle \left\langle G^{R}_{\veck_{+}}(E)G^{A}_{\veck_{-}}(E) \Im\Delta G_{\veck}(E;0,\vecq)\right\rangle \bar{\chi}(E;0,\vecq)\right\rangle_{q}
\\ \times
\frac{1 - 2a(E;0)\lambda(E;0) - \sqrt{1 - 4a(E;0)\lambda(E;0) }}{2a(E;0)^{2} \sqrt{1 - 4a(E;0)\lambda(E;0) }}\,,
\end{multline} 
\end{subequations}
where we neglected the frequency dependence of the self-energy and the CPA vertex compared to the frequency dependence of the electron-hole bubble $\chi(E;\omega,\vecq)$. 

The real and imaginary parts of the dynamical conductivity are
\begin{widetext}
\begin{subequations}\label{eq:metallic-conductivity}
\begin{align}\label{eq:metallic-conductivity-real}
\Re\sigma(E;\omega) &= \frac{e^{2}}{1 + A(E)^{2}\omega^{2}}\left[ \left\langle \Im G^{R}(E)^{2}\nabla\epsilon^{2} \right\rangle - A(E)\left( A(E)  - \frac{2\dot{\Lambda}_{I}(E;0)  \left\langle \Im G^{R}(E) \right\rangle \left\langle \Im G^{R}(E)^{3}\nabla\epsilon^{2} \right\rangle }{\Im\Sigma^{R}(E)^{2} \left\langle \Im G^{R}(E)^{2} \right\rangle}\right)\omega^{2}\right] \,,
\\ \label{eq:metallic-conductivity-imaginary}
\Im\sigma(E;\omega) &= - \frac{e^{2}}{1 + A(E)^{2}\omega^{2}}\left[A(E) \left\langle \Im G^{R}(E)^{2}\nabla\epsilon^{2} \right\rangle + \frac{2\dot{\Lambda}_{I}(E;0)  \left\langle \Im G^{R}(E) \right\rangle \left\langle \Im G^{R}(E)^{3}\nabla\epsilon^{2} \right\rangle }{\Im\Sigma^{R}(E)^{2} \left\langle \Im G^{R}(E)^{2} \right\rangle}\right]\omega
\end{align}
\end{subequations}
\end{widetext}
It is clear from the form of the dynamical conductivity that the transition to Anderson localization cannot be discerned from the static conductivity, which is standardly calculated from the Kubo formula. As discussed in Ref.~\cite{Janis:2025ab}, Anderson localization is a bound state of the propagating particle and the hole left at the origin. The static quantities are spatially averaged and do not remember the diffusing particle's initial position. Due to conservation laws, the diffusion pole of the electron-hole correlation function must be expanded to second order in frequency to observe the critical behavior of the metallic state at the Anderson localization transition.

\subsubsection{Mobility edge and instability line of the metallic phase}

The metallic phase is restricted to the region where the argument in the square root on the right-hand side of Eq.~\eqref{eq:Lambda0} is positive. In the static case, it is a real number that passes through zero at the mobility edge, as discussed in Sec.~\ref{sec:MFT-d>2}. It is a bifurcation point below which a real value of the vertex function splits into two complex conjugate roots. In analogy with classical phase transitions, a symmetry-breaking field for the Anderson localization transition, which allows one to circumvent the critical point, was proposed in Ref.~\cite{Wegner:1979aa} to be the frequency difference $\omega$ in the electron-hole correlation function $\Phi^{RA}(E;\omega,\vecq)$. However, contrary to standard statistical models of phase transitions, the frequency $\omega$ does not remove the critical point of the Anderson localization transition. The dynamic critical point at $\omega\neq 0$ does not separate a metal from an insulator, but indicates the genesis of an order parameter in the Anderson localized phase in the static limit, $\omega=0$. The critical point for $\omega \neq 0$ is determined by a change in the sign of the derivative $\partial \Im\Phi^{RA}(E;\omega,\mathbf{0})/\partial\omega$. This change occurs in the mean-field approximation when the real part of the argument of the square root in the solution for the irreducible vertex $\Lambda(E;\omega,\mathbf{0})$ goes through zero. The critical line in the $(E,\omega)$-plane from the approximation of Eq.~\eqref{eq:Lambda0} is determined by   
\begin{multline}\label{eq:Instability-line}
4\Re\left[\lambda(E_{c};\omega_{c})^{2} \left(\left\langle\chi(E_{c};\omega_{c},\vecq)^{2}\right\rangle_{\vecq}
\right.\right. \\  \left. \left. \qquad
 - \left\langle \chi(E_{c};\omega_{c},\vecq)\right\rangle_{\vecq}^{2}\right)\right] = 1 \,.
\end{multline}
This critical line separates the perturbative metallic regime from a semimetallic phase, where the perturbative expansion breaks down and the static limit cannot be reached without introducing an order parameter. The critical line $(E_{c,\omega_{c}}$ resembles the de Almeida-Thouless line in the mean-field theory of spin glasses in a magnetic field \cite{deAlmeida:1978aa}. This means that the expansion in frequency $\omega$ around the static limit cannot be used beyond the mobility edge $E_{c}$ for $\omega=0$, that is, below the instability line $(E_{c},\omega_{c})$, determined from Eq.~\eqref{eq:Instability-line}. 

We assess the behavior at the instability line for small frequencies below the mobility edge $(E_{c},0)$. Using the solution for the electron-hole irreducible vertex from Eq.~\eqref{eq:Lambda0} along the instability line, we obtain
%\
\begin{align}
\Lambda_{c}&= \frac{2\lambda_{c}\left[1 + \left(1 + i \right)\lambda_{c}\sqrt{2\Im\Delta \chi_{2}}\right]}{1 - 2\lambda_{c}\sqrt{2\Im\Delta \chi_{2}} + 4\lambda_{c}^{2}\Im\Delta \chi_{2}}
\end{align}
where we neglected the imaginary part of the mean-field vertex $\lambda_{c} = \lambda(E_{c};\omega_{c})>0$ and denoted   $\Delta\chi_{2}= \left\langle\chi(E_{c};\omega_{c},\vecq)^{2}\right\rangle_{\vecq} -  \left\langle \chi(E_{c};\omega_{c},\vecq)\right\rangle_{\vecq}^{2}$. 
%Notice that $\Im\Delta\chi_{2} \neq 0$ only below the mobility edge, along the instability line,  and $\Lambda= 2\lambda$ elsewhere. 

We use Eq.~\eqref{eq:sigmaD} to determine the dynamic conductivity along the instability line for small frequencies
\begin{align}\label{eq:dynamic-conductivity-critical}
\sigma(\omega)&= \sigma_{0}\frac{  \omega\left[\left(1 - i\alpha_{c}\right)\omega  + 2\pi n \Im\Lambda_{c} \right]}{\left(\omega + 2\pi n \Im\Lambda_{c}\right)^{2} + \alpha_{c}^{2}\omega^{2}} \,,
\end{align}
where $n$ is the particle density, $\alpha_{c} = \Im\Lambda(E_{c};\omega_{c})/ \Re\Lambda(E_{c};\omega_{c})\equiv \Im\Lambda_{c}/\Re\Lambda_{c}$, and $\sigma_{0}$ is the static conductivity at the mobility edge, calculated on the metallic side from Eq.~\eqref{eq:metallic-conductivity}. We see that $\Im\Lambda_{c}$ is the characteristic parameter of the instability line of the metallic phase and generates the real order parameter in the static limit. 

The expression for the dynamic conductivity, Eq.~\eqref{eq:dynamic-conductivity-critical}, holds only for small frequencies near the mobility edge. It qualitatively captures the differences in the behavior of the dynamic conductivity near the mobility edge $E_{c}$. The mean-field solution in the metallic phase, $\Im\Lambda(E,0)=0$, yields the Drude expression for the dynamic conductivity, with $\sigma(0) = \sigma_{0}>0$ at the mobility edge. As we approach the mobility edge along the instability line, with $\Im\Lambda_{c}\propto |\omega|^{1/{2}}$, the static conductivity vanishes with a mean-field critical exponent $s=1/2$, that is, $\sigma(\omega) \propto \sqrt{|\omega|}$. Finally, the static conductivity below the mobility edge vanishes with a critical exponent $s=1$, that is, $\sigma(\omega) \propto |\omega|$.

\subsubsection{Localized phase}

The behavior in the localized phase is more difficult to describe and less straightforward. The perturbation expansion in small frequency breaks down. The reason for this is that the microscopic irreducible vertex $\Lambda_{0}(E;\omega) = \Lambda_{R}(E;\omega) + i \Lambda_{I}(E;\omega)$ has a discontinuity in its imaginary part at zero frequency, $\Lambda_{I}(E;0_{\pm})= \pm |\Lambda_{I}(E;0)|$. This jump is transferred to the conserving vertex and the electron-hole correlation function.    
In the localized phase, its denominator in the mean-field approximation has a more complex structure than assumed in Ref.~\cite{Janis:2025ab} because of the nontrivial structure of the bound electron-hole pair 
\begin{multline}
\mathcal{D}(E;\omega,\mathbf{q}) \doteq  \frac{1 + i\alpha_{0}(E)}{2| \Im \Sigma^{R}(E)|}
\\ \times
\left[ - i\left( \omega  + \frac{2\alpha_{0}(E)|\Im\Sigma^{R}(E)|}{1 + \alpha_{0}(E)^{2}}\right) +\ D(E)q^{2} 
\right. \\ \left.
 -\ \frac{2\alpha_{0}(E)^{2}|\Im\Sigma^{R}(E)|}{1 + \alpha_{0}(E)^{2}}\right] \,,
\end{multline}
with $\alpha_{0}(E) = \Lambda_{I}(E;0)/\Lambda_{R}(E;0)$. We used the mean-field diffusion constant 
\be
D(E) = \frac{\left\langle \Im G^{R}(E)^{2} \nabla \epsilon^{2}\right\rangle}{\left|\left\langle \Im G^{R}(E) \right\rangle \right|}\,.
\ee
%
%We recall that $\alpha_{0}  |\alpha_{0}|\mathrm{sign}(\omega)$.
%\be
%\nabla_{\vecq}^{2} \mathcal{D}(E;0,\mathbf{q}) \doteq \left(\Lambda_{R} + i\Lambda_{I}\right)\left\langle \nabla G^{R}\nabla G^{A} \right\rangle
%\ee

%\be
%\sigma(\omega,\vecq) = \frac{- i e^{2}\omega D\left(1 + \alpha_{0}^{2}\right) }{- i\left(\omega + 2\alpha_{0}|\Im\Sigma|\right)  +  \alpha_{0}\left(\omega - 2\alpha_{0}|\Im\Sigma|\right) + \left(1 + \alpha_{0}^{2}\right)Dq^{2}}    
%\ee

The terms proportional to $\alpha_{0}$ cause the breakdown of linear response theory, and Eq.~\eqref{eq:sigmaD} for the dynamical conductivity cannot be used. Instead, the time propagation in the electron-hole correlation function should be analyzed at long distances, in the limit $q\to 0$. It is easy to see that there is no long-range diffusion if 
\be
D(E)q^{2} - \frac{2\alpha_{0}(E)^{2}|\Im\Sigma^{R}(E)|}{1 + \alpha_{0}(E)^{2}} <0\,.
\ee  
Diffusion can be restored if the particle acquires sufficient energy to break the bond of the electron-hole pair. The threshold wave vector $q_{0}$ above which diffusion is restored in the localized phase is then 
\begin{align}
q_{0}^{2} &= \frac{2|\Im\Sigma^{R}(E)|\alpha_{0}(E)^{2} }{D(E)\left(1 + \alpha_{0}(E)^{2}\right)} \,.
%\\
%\xi^{2} &= \frac{D}{2|\Im\Sigma|}\left(1 + \frac i{\alpha_{0}}\right)
\end{align}
The energy of the localized bound state equals the kinetic energy of the particle with the threshold wave vector
\be
E_{b}(E) = \frac{\hbar^{2}|\Im\Sigma^{R}(E)|\alpha_{0}(E)^{2}}{mD(E) \left(1 + \alpha_{0}(E)^{2}\right)}\,,
\ee
where $m$ is the particle's mass. The threshold energy is given by the applied electric field $U_{b} = E_{b}/|e|$. The electric current flows in the direction of the electric field when $U > U_{b}$, and the system becomes metallic. 

The localized particle in the static limit, $\omega=0$, is characterized by a complex wave vector $q= q_{R} + iq_{I}$ with the following components
\begin{subequations}
\begin{multline}
|q_{R}| = \sqrt{\frac{|\Im\Sigma^{R}(E)\alpha_{0}(E)|}{D(E)\left(1 + \alpha_{0}(E)^{2}\right)}}
\\ \times
\frac 1{\sqrt{\sqrt{1 + \alpha_{0}(E)^{2}} - |\alpha_{0}(E)|}} \,,
\end{multline}
\begin{multline}
|q_{I}| = \sqrt{\frac{|\Im\Sigma^{R}(E)\alpha_{0}(E)|}{D(E)\left(1 + \alpha_{0}(E)^{2}\right)}}
\\ \times
\sqrt{\sqrt{1 + \alpha_{0}(E)^{2}} - |\alpha_{0}(E)|}\,. 
\end{multline}
\end{subequations}
The real component causes modulation of the phase of the wave function of the electron-hole bound pair in the long-time limit $t\to\infty$. The imaginary component determines the localization length of the particle density decay from origin  $|n(x,t \to\infty)| = n(0,0) e^{-|x|/\xi}$. The localization length is 
\begin{multline}
\xi(E) =  \sqrt{\frac{D(E)\left(1 + \alpha_{0}(E)^{2}\right)}{|\Im\Sigma^{R}(E)\alpha_{0}(E)|}}
\\ \times
\frac {\pi}{\sqrt{\sqrt{1 + \alpha_{0}(E)^{2}} - |\alpha_{0}|}} \,.
\end{multline}
It diverges at the Anderson localization transition as $|\alpha_{0}(E)|^{-1/2}$ and matches the divergent scale $A(E)$ from the metallic side.

\section{Conclusions}

There are various options and effective statistical models for simulating and studying the critical behavior of the Anderson localization transition. However, to understand the microscopic origin of the vanishing of charge diffusion and the emergence of Anderson localization, it is necessary to analyze the original model of a quantum particle propagating in a random environment. The fundamental quantities to be studied in the thermodynamic limit are the configurationally averaged one-particle and two-particle resolvents. The microscopic quantities fully characterizing them are the one-particle self-energy and the two-particle irreducible vertices. These functions must evidence the Anderson localization transition and the phases on both sides of the transition. 

We presented a general scheme for constructing approximations in renormalized perturbation theory that interpolate between the weak- and strong-disorder limits, including the Anderson localization transition. There are three principal aspects that make the approximations successful. First, the two-particle vertex must capture more of the disorder's impact on particle propagation than the one-particle self-energy. Second, the microscopic dynamical quantities must be separated from the macroscopic observables that obey conservation laws, because causality of the self-energy and the two-particle vertices is incompatible with the dynamical Ward identities. Third, local and nonlocal scattering processes must be treated separately, as the two-particle self-consistency required for Anderson localization can be achieved only with nonlocal, off-diagonal one-particle propagators.           

Perturbation theory uses two-particle-irreducible vertices as its fundamental ingredients to provide a more detailed picture than the self-energy. The consistency between the one-particle and two-particle Green functions is ensured by matching the imaginary part of the self-energy with the real part of the static and homogeneous electron-hole irreducible vertex via a Ward identity, Eq.~\eqref{eq:KK-nonlocal}. The complete dynamical Ward identity cannot be imposed on approximations for the dynamical two-particle vertices. The irreducible conserving two-particle vertex, $L^{RA}_{\veck\veck'}(E;\omega,\vecq)$, is recovered from the dynamical one, $\Lambda^{RA}_{\veck\veck'}(E;\omega,\vecq)$, via a given relation, Eqs.~\eqref{eq:L-Lambda}, \eqref{eq:R-general}, ensuring that the conservation laws of the response functions and macroscopic observables are satisfied.  

The contributions to the two-particle irreducible vertices must be split into local and nonlocal terms. The local contributions are summed using the coherent potential approximation. Its local vertex serves as input for the expansion beyond this solution to obtain nontrivial two-particle vertex functions. Nonlocal scattering allows one to distinguish between electron-electron (cooperon) and electron-hole (diffuson) multiple scattering. Their self-consistent interconnection yields two-channel parquet equations, nonlinear integral equations with a bifurcation critical point for strong disorder, representing the Anderson localization transition. 

We employed a local, mean-field approximation to the parquet equations' integral kernel to obtain quantitative results for the behavior of the conserving electron-hole vertex, the electron-hole correlation function, and the dynamical conductivity in both metallic and localized phases. We observed a mean-field critical behavior at the Anderson localization transition and the genesis of an order parameter in the localized phase. We calculated mean-field characteristics of an electron propagating in the localized phase, spatially restricted by a quantum bound state with a hole left at the origin. We assessed the localization length and the energy barrier required to drive an electric current in the localized phase.    

The presented construction provides a microscopic explanation of Anderson localization as the emergence of a quantum bound state between the propagating particle and the hole left behind. The existence of a bound state is signaled by a gap in the dynamical conductivity, corresponding to the threshold energy required to generate an electric current and turn the electron's localized wave function into an extended one. This result can be generalized to conclude that a gap in a two-particle response function, without a gap in the self-energy, indicates the existence of quantum bound states without their condensation. Linear response theory breaks down beyond the localization transition, and static quantities with no memory, such as conductivity and the diffusion constant, cannot describe particle propagation in the Anderson-localized phase. Only the fully dynamical response functions contain imprints of Anderson localization. 

We conclude that, in this paper, we have completed the construction of a consistent mean-field theory of a new phase in disordered electron systems, extending Anderson localization to dynamical quantities. We found that the new phase emerges bekow an instability line in the $(E,\omega)$-plane, where the frequency derivative of the homogeneous electron-hole correlation function, $\partial \Im\Phi^{RA}(E;\omega,\mathbf{0})/\partial\omega$ from Eq.~\eqref{eq:Phi-AR}, changes sign from positive above the line (normal metallic phase) to negative below it (semimetallic phase). The best quantity to check this behavior is the electric polarizability.       

% \end{widetext}

%\bibliographystyle{apsrev}
%\bibliography{/Users/vaclav/Dropbox/TeX/BibTeX/parquets_MB,/Users/vaclav/Dropbox/TeX/BibTeX/localization_anderson,/Users/vaclav/Dropbox/TeX/BibTeX/Spin-glasses1}

\begin{thebibliography}{54}%
\makeatletter
\providecommand \@ifxundefined [1]{%
 \@ifx{#1\undefined}
}%
\providecommand \@ifnum [1]{%
 \ifnum #1\expandafter \@firstoftwo
 \else \expandafter \@secondoftwo
 \fi
}%
\providecommand \@ifx [1]{%
 \ifx #1\expandafter \@firstoftwo
 \else \expandafter \@secondoftwo
 \fi
}%
\providecommand \natexlab [1]{#1}%
\providecommand \enquote  [1]{``#1''}%
\providecommand \bibnamefont  [1]{#1}%
\providecommand \bibfnamefont [1]{#1}%
\providecommand \citenamefont [1]{#1}%
\providecommand \href@noop [0]{\@secondoftwo}%
\providecommand \href [0]{\begingroup \@sanitize@url \@href}%
\providecommand \@href[1]{\@@startlink{#1}\@@href}%
\providecommand \@@href[1]{\endgroup#1\@@endlink}%
\providecommand \@sanitize@url [0]{\catcode `\\12\catcode `\$12\catcode
  `\&12\catcode `\#12\catcode `\^12\catcode `\_12\catcode `\%12\relax}%
\providecommand \@@startlink[1]{}%
\providecommand \@@endlink[0]{}%
\providecommand \url  [0]{\begingroup\@sanitize@url \@url }%
\providecommand \@url [1]{\endgroup\@href {#1}{\urlprefix }}%
\providecommand \urlprefix  [0]{URL }%
\providecommand \Eprint [0]{\href }%
\providecommand \doibase [0]{https://doi.org/}%
\providecommand \selectlanguage [0]{\@gobble}%
\providecommand \bibinfo  [0]{\@secondoftwo}%
\providecommand \bibfield  [0]{\@secondoftwo}%
\providecommand \translation [1]{[#1]}%
\providecommand \BibitemOpen [0]{}%
\providecommand \bibitemStop [0]{}%
\providecommand \bibitemNoStop [0]{.\EOS\space}%
\providecommand \EOS [0]{\spacefactor3000\relax}%
\providecommand \BibitemShut  [1]{\csname bibitem#1\endcsname}%
\let\auto@bib@innerbib\@empty
%</preamble>
\bibitem [{\citenamefont {Jani{\v s}}(2025)}]{Janis:2025ab}%
  \BibitemOpen
  \bibfield  {author} {\bibinfo {author} {\bibfnamefont {V.}~\bibnamefont
  {Jani{\v s}}},\ }\bibfield  {title} {\bibinfo {title} {Anderson localization:
  a disorder-induced quantum bound state},\ }\href
  {https://doi.org/10.1088/1367-2630/adea17} {\bibfield  {journal} {\bibinfo
  {journal} {New Journal of Physics}\ }\textbf {\bibinfo {volume} {27}},\
  \bibinfo {pages} {073503} (\bibinfo {year} {2025})}\BibitemShut {NoStop}%
\bibitem [{\citenamefont {Anderson}(1958)}]{Anderson:1958aa}%
  \BibitemOpen
  \bibfield  {author} {\bibinfo {author} {\bibfnamefont {P.~W.}\ \bibnamefont
  {Anderson}},\ }\bibfield  {title} {\bibinfo {title} {Absence of diffusion in
  certain random lattices},\ }\href {https://doi.org/10.1103/PhysRev.109.1492}
  {\bibfield  {journal} {\bibinfo  {journal} {Phys. Rev.}\ }\textbf {\bibinfo
  {volume} {109}},\ \bibinfo {pages} {1492} (\bibinfo {year}
  {1958})}\BibitemShut {NoStop}%
\bibitem [{\citenamefont {Mott}\ and\ \citenamefont
  {Twose}(1961)}]{Mott:1961aa}%
  \BibitemOpen
  \bibfield  {author} {\bibinfo {author} {\bibfnamefont {N.}~\bibnamefont
  {Mott}}\ and\ \bibinfo {author} {\bibfnamefont {W.}~\bibnamefont {Twose}},\
  }\bibfield  {title} {\bibinfo {title} {The theory of impurity conduction},\
  }\href {https://doi.org/10.1080/00018736100101271} {\bibfield  {journal}
  {\bibinfo  {journal} {Advances in Physics}\ }\textbf {\bibinfo {volume}
  {10}},\ \bibinfo {pages} {107} (\bibinfo {year} {1961})}\BibitemShut
  {NoStop}%
\bibitem [{\citenamefont {Mott}(1967)}]{Mott:1967aa}%
  \BibitemOpen
  \bibfield  {author} {\bibinfo {author} {\bibfnamefont {N.}~\bibnamefont
  {Mott}},\ }\bibfield  {title} {\bibinfo {title} {Electrons in disordered
  structures},\ }\href {https://doi.org/10.1080/00018736700101265} {\bibfield
  {journal} {\bibinfo  {journal} {Advances in Physics}\ }\textbf {\bibinfo
  {volume} {16}},\ \bibinfo {pages} {49} (\bibinfo {year} {1967})}\BibitemShut
  {NoStop}%
\bibitem [{\citenamefont {Mott}(1970)}]{Mott:1970aa}%
  \BibitemOpen
  \bibfield  {author} {\bibinfo {author} {\bibfnamefont {N.~F.}\ \bibnamefont
  {Mott}},\ }\bibfield  {title} {\bibinfo {title} {Conduction in
  non-crystalline systems: Iv. anderson localization in a disordered lattice},\
  }\href {https://doi.org/10.1080/14786437008228147} {\bibfield  {journal}
  {\bibinfo  {journal} {Philosophical Magazine}\ }\textbf {\bibinfo {volume}
  {22}},\ \bibinfo {pages} {7} (\bibinfo {year} {1970})}\BibitemShut {NoStop}%
\bibitem [{\citenamefont {Thouless}(1970)}]{Thouless:1970aa}%
  \BibitemOpen
  \bibfield  {author} {\bibinfo {author} {\bibfnamefont {D.~J.}\ \bibnamefont
  {Thouless}},\ }\bibfield  {title} {\bibinfo {title} {Anderson's theory of
  localized states},\ }\href@noop {} {\bibfield  {journal} {\bibinfo  {journal}
  {Journal of Physics C: Solid State Physics}\ }\textbf {\bibinfo {volume}
  {3}},\ \bibinfo {pages} {1559 } (\bibinfo {year} {1970})}\BibitemShut
  {NoStop}%
\bibitem [{\citenamefont {Abou-Chacra}\ \emph {et~al.}(1973)\citenamefont
  {Abou-Chacra}, \citenamefont {Thouless},\ and\ \citenamefont
  {Anderson}}]{Abou-Chacra:1973aa}%
  \BibitemOpen
  \bibfield  {author} {\bibinfo {author} {\bibfnamefont {R.}~\bibnamefont
  {Abou-Chacra}}, \bibinfo {author} {\bibfnamefont {D.~J.}\ \bibnamefont
  {Thouless}},\ and\ \bibinfo {author} {\bibfnamefont {P.~W.}\ \bibnamefont
  {Anderson}},\ }\bibfield  {title} {\bibinfo {title} {A selfconsistent theory
  of localization},\ }\href {https://doi.org/10.1088/0022-3719/6/10/009}
  {\bibfield  {journal} {\bibinfo  {journal} {Journal of Physics C: Solid State
  Physics}\ }\textbf {\bibinfo {volume} {6}},\ \bibinfo {pages} {1734}
  (\bibinfo {year} {1973})}\BibitemShut {NoStop}%
\bibitem [{\citenamefont {Thouless}(1974)}]{Thouless:1974aa}%
  \BibitemOpen
  \bibfield  {author} {\bibinfo {author} {\bibfnamefont {D.~J.}\ \bibnamefont
  {Thouless}},\ }\bibfield  {title} {\bibinfo {title} {Electrons in disordered
  systems and the theory of localization},\ }\href
  {https://doi.org/10.1016/0370-1573(74)90029-5} {\bibfield  {journal}
  {\bibinfo  {journal} {Physics Reports}\ }\textbf {\bibinfo {volume} {13}},\
  \bibinfo {pages} {93} (\bibinfo {year} {1974})}\BibitemShut {NoStop}%
\bibitem [{\citenamefont {Licciardello}\ and\ \citenamefont
  {Thouless}(1975)}]{Licciardello:1975aa}%
  \BibitemOpen
  \bibfield  {author} {\bibinfo {author} {\bibfnamefont {D.~C.}\ \bibnamefont
  {Licciardello}}\ and\ \bibinfo {author} {\bibfnamefont {D.~J.}\ \bibnamefont
  {Thouless}},\ }\bibfield  {title} {\bibinfo {title} {Constancy of minimum
  metallic conductivity in two dimensions},\ }\href
  {https://doi.org/10.1103/PhysRevLett.35.1475} {\bibfield  {journal} {\bibinfo
   {journal} {Physical Review Letters}\ }\textbf {\bibinfo {volume} {35}},\
  \bibinfo {pages} {1475} (\bibinfo {year} {1975})}\BibitemShut {NoStop}%
\bibitem [{\citenamefont {Wegner}(1976)}]{Wegner:1976aa}%
  \BibitemOpen
  \bibfield  {author} {\bibinfo {author} {\bibfnamefont {F.~J.}\ \bibnamefont
  {Wegner}},\ }\bibfield  {title} {\bibinfo {title} {Electrons in disordered
  systems. scaling near the mobility edge},\ }\href
  {https://doi.org/10.1007/BF01315248} {\bibfield  {journal} {\bibinfo
  {journal} {Zeitschrift f{\"u}r Physik B Condensed Matter}\ }\textbf {\bibinfo
  {volume} {25}},\ \bibinfo {pages} {327} (\bibinfo {year} {1976})}\BibitemShut
  {NoStop}%
\bibitem [{\citenamefont {Wegner}(1979)}]{Wegner:1979aa}%
  \BibitemOpen
  \bibfield  {author} {\bibinfo {author} {\bibfnamefont {F.}~\bibnamefont
  {Wegner}},\ }\bibfield  {title} {\bibinfo {title} {The mobility edge problem:
  Continuous symmetry and a conjecture},\ }\href
  {https://doi.org/10.1007/BF01319839} {\bibfield  {journal} {\bibinfo
  {journal} {Zeitschrift f{\"u}r Physik B Condensed Matter}\ }\textbf {\bibinfo
  {volume} {35}},\ \bibinfo {pages} {207} (\bibinfo {year} {1979})}\BibitemShut
  {NoStop}%
\bibitem [{\citenamefont {Hikami}(1981)}]{Hikami:1981aa}%
  \BibitemOpen
  \bibfield  {author} {\bibinfo {author} {\bibfnamefont {S.}~\bibnamefont
  {Hikami}},\ }\bibfield  {title} {\bibinfo {title} {Anderson localization in a
  nonlinear-sigma-model reprentation},\ }\href
  {https://doi.org/10.1103/PhysRevB.24.2671} {\bibfield  {journal} {\bibinfo
  {journal} {Physical Review B}\ }\textbf {\bibinfo {volume} {24}},\ \bibinfo
  {pages} {2671} (\bibinfo {year} {1981})}\BibitemShut {NoStop}%
\bibitem [{\citenamefont {Abrahams}\ \emph {et~al.}(1979)\citenamefont
  {Abrahams}, \citenamefont {Anderson}, \citenamefont {Licciardello},\ and\
  \citenamefont {Ramakrishnan}}]{Abrahams:1979aa}%
  \BibitemOpen
  \bibfield  {author} {\bibinfo {author} {\bibfnamefont {E.}~\bibnamefont
  {Abrahams}}, \bibinfo {author} {\bibfnamefont {P.~W.}\ \bibnamefont
  {Anderson}}, \bibinfo {author} {\bibfnamefont {D.~C.}\ \bibnamefont
  {Licciardello}},\ and\ \bibinfo {author} {\bibfnamefont {T.~V.}\ \bibnamefont
  {Ramakrishnan}},\ }\bibfield  {title} {\bibinfo {title} {Scaling theory of
  localization: Absence of quantum diffusion in two dimensions},\ }\href
  {https://doi.org/10.1103/PhysRevLett.42.673} {\bibfield  {journal} {\bibinfo
  {journal} {Physical Review Letters}\ }\textbf {\bibinfo {volume} {42}},\
  \bibinfo {pages} {673} (\bibinfo {year} {1979})}\BibitemShut {NoStop}%
\bibitem [{\citenamefont {Evers}\ and\ \citenamefont
  {Mirlin}(2008)}]{Evers:2008aa}%
  \BibitemOpen
  \bibfield  {author} {\bibinfo {author} {\bibfnamefont {F.}~\bibnamefont
  {Evers}}\ and\ \bibinfo {author} {\bibfnamefont {A.}~\bibnamefont {Mirlin}},\
  }\bibfield  {title} {\bibinfo {title} {Anderson transitions},\ }\href
  {https://doi.org/10.1103/RevModPhys.80.1355} {\bibfield  {journal} {\bibinfo
  {journal} {Reviews of Modern Physics}\ }\textbf {\bibinfo {volume} {80}},\
  \bibinfo {pages} {1355} (\bibinfo {year} {2008})}\BibitemShut {NoStop}%
\bibitem [{\citenamefont {Efetov}(1987)}]{Efetov:1987aa}%
  \BibitemOpen
  \bibfield  {author} {\bibinfo {author} {\bibfnamefont {K.~B.}\ \bibnamefont
  {Efetov}},\ }\bibfield  {title} {\bibinfo {title} {Density-density correlator
  in a model of a disordered metal on a bethe lattice},\ }\href@noop {}
  {\bibfield  {journal} {\bibinfo  {journal} {Zh. Exp. Teor. Fiz.}\ }\textbf
  {\bibinfo {volume} {92}},\ \bibinfo {pages} {638} (\bibinfo {year}
  {1987})}\BibitemShut {NoStop}%
\bibitem [{\citenamefont {Zirnbauer}(1986)}]{Zirnbauer:1986aa}%
  \BibitemOpen
  \bibfield  {author} {\bibinfo {author} {\bibfnamefont {M.~R.}\ \bibnamefont
  {Zirnbauer}},\ }\bibfield  {title} {\bibinfo {title} {Anderson localization
  and non-linear sigma model with graded symmetry},\ }\href
  {https://doi.org/https://doi.org/10.1016/0550-3213(86)90316-0} {\bibfield
  {journal} {\bibinfo  {journal} {Nuclear Physics B}\ }\textbf {\bibinfo
  {volume} {265}},\ \bibinfo {pages} {375} (\bibinfo {year}
  {1986})}\BibitemShut {NoStop}%
\bibitem [{\citenamefont {Mirlin}\ and\ \citenamefont
  {Fyodorov}(1994)}]{Mirlin:1994aa}%
  \BibitemOpen
  \bibfield  {author} {\bibinfo {author} {\bibfnamefont {A.~D.}\ \bibnamefont
  {Mirlin}}\ and\ \bibinfo {author} {\bibfnamefont {Y.~V.}\ \bibnamefont
  {Fyodorov}},\ }\bibfield  {title} {\bibinfo {title} {Distribution of local
  densities of states, order parameter function, and critical behavior near the
  anderson transition},\ }\href {https://doi.org/10.1103/PhysRevLett.72.526}
  {\bibfield  {journal} {\bibinfo  {journal} {Physical Review Letters}\
  }\textbf {\bibinfo {volume} {72}},\ \bibinfo {pages} {526} (\bibinfo {year}
  {1994})}\BibitemShut {NoStop}%
\bibitem [{\citenamefont {Tarquini}\ \emph {et~al.}(2017)\citenamefont
  {Tarquini}, \citenamefont {Biroli},\ and\ \citenamefont
  {Tarzia}}]{Tarquini:2017aa}%
  \BibitemOpen
  \bibfield  {author} {\bibinfo {author} {\bibfnamefont {E.}~\bibnamefont
  {Tarquini}}, \bibinfo {author} {\bibfnamefont {G.}~\bibnamefont {Biroli}},\
  and\ \bibinfo {author} {\bibfnamefont {M.}~\bibnamefont {Tarzia}},\
  }\bibfield  {title} {\bibinfo {title} {Critical properties of the anderson
  localization transition and the high-dimensional limit},\ }\href
  {https://doi.org/10.1103/PhysRevB.95.094204} {\bibfield  {journal} {\bibinfo
  {journal} {Physical Review B}\ }\textbf {\bibinfo {volume} {95}},\ \bibinfo
  {pages} {094204} (\bibinfo {year} {2017})}\BibitemShut {NoStop}%
\bibitem [{\citenamefont {Parisi}\ \emph {et~al.}(2019)\citenamefont {Parisi},
  \citenamefont {Pascazio}, \citenamefont {Pietracaprina}, \citenamefont
  {Ros},\ and\ \citenamefont {Scardicchio}}]{Parisi:2020aa}%
  \BibitemOpen
  \bibfield  {author} {\bibinfo {author} {\bibfnamefont {G.}~\bibnamefont
  {Parisi}}, \bibinfo {author} {\bibfnamefont {S.}~\bibnamefont {Pascazio}},
  \bibinfo {author} {\bibfnamefont {F.}~\bibnamefont {Pietracaprina}}, \bibinfo
  {author} {\bibfnamefont {V.}~\bibnamefont {Ros}},\ and\ \bibinfo {author}
  {\bibfnamefont {A.}~\bibnamefont {Scardicchio}},\ }\bibfield  {title}
  {\bibinfo {title} {Anderson transition on the bethe lattice: an approach with
  real energies},\ }\href {https://doi.org/10.1088/1751-8121/ab56e8} {\bibfield
   {journal} {\bibinfo  {journal} {Journal of Physics A: Mathematical and
  Theoretical}\ }\textbf {\bibinfo {volume} {53}},\ \bibinfo {pages} {014003}
  (\bibinfo {year} {2019})}\BibitemShut {NoStop}%
\bibitem [{\citenamefont {Bollob{\'a}s}(1998)}]{Bollobas:1998aa}%
  \BibitemOpen
  \bibfield  {author} {\bibinfo {author} {\bibfnamefont {B.}~\bibnamefont
  {Bollob{\'a}s}},\ }\bibinfo {title} {Random graphs},\ in\ \href@noop {}
  {\emph {\bibinfo {booktitle} {Modern Graph Theory}}}\ (\bibinfo  {publisher}
  {Springer New York},\ \bibinfo {year} {1998})\ pp.\ \bibinfo {pages}
  {215--252}\BibitemShut {NoStop}%
\bibitem [{\citenamefont {Garc{\'\i}a-Mata}\ \emph {et~al.}(2017)\citenamefont
  {Garc{\'\i}a-Mata}, \citenamefont {Martin}, \citenamefont {Giraud},
  \citenamefont {Georgeot}, \citenamefont {Dubertrand},\ and\ \citenamefont
  {Lemeri{\'e}}}]{Garcia-Mata:2017aa}%
  \BibitemOpen
  \bibfield  {author} {\bibinfo {author} {\bibfnamefont {I.}~\bibnamefont
  {Garc{\'\i}a-Mata}}, \bibinfo {author} {\bibfnamefont {J.}~\bibnamefont
  {Martin}}, \bibinfo {author} {\bibfnamefont {O.}~\bibnamefont {Giraud}},
  \bibinfo {author} {\bibfnamefont {B.}~\bibnamefont {Georgeot}}, \bibinfo
  {author} {\bibfnamefont {R.}~\bibnamefont {Dubertrand}},\ and\ \bibinfo
  {author} {\bibfnamefont {G.}~\bibnamefont {Lemeri{\'e}}},\ }\bibfield
  {title} {\bibinfo {title} {Scaling theory of the anderson transition in
  random graphs: Ergodicity and universality},\ }\href
  {https://doi.org/10.1103/PhysRevLett.118.166801} {\bibfield  {journal}
  {\bibinfo  {journal} {Physical Review Letters}\ }\textbf {\bibinfo {volume}
  {118}},\ \bibinfo {pages} {166801} (\bibinfo {year} {2017})}\BibitemShut
  {NoStop}%
\bibitem [{\citenamefont {Tikhonov}\ and\ \citenamefont
  {Mirlin}(2019)}]{Tikhonov:2019aa}%
  \BibitemOpen
  \bibfield  {author} {\bibinfo {author} {\bibfnamefont {K.~S.}\ \bibnamefont
  {Tikhonov}}\ and\ \bibinfo {author} {\bibfnamefont {A.~D.}\ \bibnamefont
  {Mirlin}},\ }\bibfield  {title} {\bibinfo {title} {Critical behavior at the
  localization transition on random regular graphs},\ }\href
  {https://doi.org/10.1103/PhysRevB.99.214202} {\bibfield  {journal} {\bibinfo
  {journal} {Physical Review B}\ }\textbf {\bibinfo {volume} {99}},\ \bibinfo
  {pages} {214202} (\bibinfo {year} {2019})}\BibitemShut {NoStop}%
\bibitem [{\citenamefont {Garc{\'\i}a-Mata}\ \emph {et~al.}(2020)\citenamefont
  {Garc{\'\i}a-Mata}, \citenamefont {Martin}, \citenamefont {Dubertrand},
  \citenamefont {Giraud}, \citenamefont {Georgeot},\ and\ \citenamefont
  {Lemeri{\'e}}}]{Garcia-Mata:2020aa}%
  \BibitemOpen
  \bibfield  {author} {\bibinfo {author} {\bibfnamefont {I.}~\bibnamefont
  {Garc{\'\i}a-Mata}}, \bibinfo {author} {\bibfnamefont {J.}~\bibnamefont
  {Martin}}, \bibinfo {author} {\bibfnamefont {R.}~\bibnamefont {Dubertrand}},
  \bibinfo {author} {\bibfnamefont {O.}~\bibnamefont {Giraud}}, \bibinfo
  {author} {\bibfnamefont {B.}~\bibnamefont {Georgeot}},\ and\ \bibinfo
  {author} {\bibfnamefont {G.}~\bibnamefont {Lemeri{\'e}}},\ }\bibfield
  {title} {\bibinfo {title} {Two critical localization lengths in the anderson
  transition on random graphs},\ }\href
  {https://doi.org/10.1103/PhysRevResearch.2.012020} {\bibfield  {journal}
  {\bibinfo  {journal} {Physical Review Research}\ }\textbf {\bibinfo {volume}
  {2}},\ \bibinfo {pages} {012020} (\bibinfo {year} {2020})}\BibitemShut
  {NoStop}%
\bibitem [{\citenamefont {Garc{\'\i}a-Mata}\ \emph {et~al.}(2022)\citenamefont
  {Garc{\'\i}a-Mata}, \citenamefont {Martin}, \citenamefont {Giraud},
  \citenamefont {Georgeot}, \citenamefont {Dubertrand},\ and\ \citenamefont
  {Lemeri{\'e}}}]{Garcia-Mata:2022aa}%
  \BibitemOpen
  \bibfield  {author} {\bibinfo {author} {\bibfnamefont {I.}~\bibnamefont
  {Garc{\'\i}a-Mata}}, \bibinfo {author} {\bibfnamefont {J.}~\bibnamefont
  {Martin}}, \bibinfo {author} {\bibfnamefont {O.}~\bibnamefont {Giraud}},
  \bibinfo {author} {\bibfnamefont {B.}~\bibnamefont {Georgeot}}, \bibinfo
  {author} {\bibfnamefont {R.}~\bibnamefont {Dubertrand}},\ and\ \bibinfo
  {author} {\bibfnamefont {G.}~\bibnamefont {Lemeri{\'e}}},\ }\bibfield
  {title} {\bibinfo {title} {Critical properties of the anderson transition on
  random graphs: Two-parameter scaling theory, kosterlitz-thouless type flow,
  and many-body localization},\ }\href
  {https://doi.org/10.1103/PhysRevB.106.214202} {\bibfield  {journal} {\bibinfo
   {journal} {Physical Review B}\ }\textbf {\bibinfo {volume} {106}},\ \bibinfo
  {pages} {214202} (\bibinfo {year} {2022})}\BibitemShut {NoStop}%
\bibitem [{\citenamefont {Mirlin}(2000)}]{Mirlin:2000aa}%
  \BibitemOpen
  \bibfield  {author} {\bibinfo {author} {\bibfnamefont {A.~D.}\ \bibnamefont
  {Mirlin}},\ }\bibfield  {title} {\bibinfo {title} {Statistics of energy
  levels and eigenfunctions in disordered systems},\ }\href
  {https://doi.org/https://doi.org/10.1016/S0370-1573(99)00091-5} {\bibfield
  {journal} {\bibinfo  {journal} {Physics Reports}\ }\textbf {\bibinfo {volume}
  {326}},\ \bibinfo {pages} {259} (\bibinfo {year} {2000})}\BibitemShut
  {NoStop}%
\bibitem [{\citenamefont {Fr{\"o}hlich}\ and\ \citenamefont
  {Spencer}(1983)}]{Frohlich:1983aa}%
  \BibitemOpen
  \bibfield  {author} {\bibinfo {author} {\bibfnamefont {J.}~\bibnamefont
  {Fr{\"o}hlich}}\ and\ \bibinfo {author} {\bibfnamefont {T.}~\bibnamefont
  {Spencer}},\ }\bibfield  {title} {\bibinfo {title} {Absence of diffusion in
  the anderson tight abdence of diffusion in the anderson tight binding model
  for large disorder or low energy},\ }\href@noop {} {\bibfield  {journal}
  {\bibinfo  {journal} {Communications in Mathematical Physics}\ }\textbf
  {\bibinfo {volume} {88}},\ \bibinfo {pages} {151} (\bibinfo {year}
  {1983})}\BibitemShut {NoStop}%
\bibitem [{\citenamefont {Kramer}\ and\ \citenamefont
  {MacKinnon}(1993)}]{Kramer:1993aa}%
  \BibitemOpen
  \bibfield  {author} {\bibinfo {author} {\bibfnamefont {B.}~\bibnamefont
  {Kramer}}\ and\ \bibinfo {author} {\bibfnamefont {A.}~\bibnamefont
  {MacKinnon}},\ }\bibfield  {title} {\bibinfo {title} {Localization: theory
  and experiment},\ }\href {https://doi.org/10.1088/0034-4885/56/12/001}
  {\bibfield  {journal} {\bibinfo  {journal} {Rep. Prog. Phys.}\ }\textbf
  {\bibinfo {volume} {56}},\ \bibinfo {pages} {1469} (\bibinfo {year}
  {1993})}\BibitemShut {NoStop}%
\bibitem [{\citenamefont {{Marko\v s}}(2006)}]{Markos:2006aa}%
  \BibitemOpen
  \bibfield  {author} {\bibinfo {author} {\bibfnamefont {P.}~\bibnamefont
  {{Marko\v s}}},\ }\bibfield  {title} {\bibinfo {title} {Numerical analysis of
  the anderson localization},\ }\href@noop {} {\bibfield  {journal} {\bibinfo
  {journal} {Acta Phys. Slovaca}\ }\textbf {\bibinfo {volume} {56}},\ \bibinfo
  {pages} {561} (\bibinfo {year} {2006})}\BibitemShut {NoStop}%
\bibitem [{\citenamefont {Segev}\ \emph {et~al.}(2013)\citenamefont {Segev},
  \citenamefont {Silberberg},\ and\ \citenamefont
  {Christodoulides}}]{Segev:2013aa}%
  \BibitemOpen
  \bibfield  {author} {\bibinfo {author} {\bibfnamefont {M.}~\bibnamefont
  {Segev}}, \bibinfo {author} {\bibfnamefont {Y.}~\bibnamefont {Silberberg}},\
  and\ \bibinfo {author} {\bibfnamefont {D.~N.}\ \bibnamefont
  {Christodoulides}},\ }\bibfield  {title} {\bibinfo {title} {Anderson
  localization of light},\ }\href {https://doi.org/10.1038/nphoton.2013.30}
  {\bibfield  {journal} {\bibinfo  {journal} {Nature Photonics}\ }\textbf
  {\bibinfo {volume} {7}},\ \bibinfo {pages} {197} (\bibinfo {year}
  {2013})}\BibitemShut {NoStop}%
\bibitem [{\citenamefont {Yamilov}\ \emph {et~al.}(2023)\citenamefont
  {Yamilov}, \citenamefont {Skipetrov}, \citenamefont {Hughes}, \citenamefont
  {Minkov}, \citenamefont {Yu},\ and\ \citenamefont {Cao}}]{Yamilov:2023aa}%
  \BibitemOpen
  \bibfield  {author} {\bibinfo {author} {\bibfnamefont {A.}~\bibnamefont
  {Yamilov}}, \bibinfo {author} {\bibfnamefont {S.~E.}\ \bibnamefont
  {Skipetrov}}, \bibinfo {author} {\bibfnamefont {T.~W.}\ \bibnamefont
  {Hughes}}, \bibinfo {author} {\bibfnamefont {M.}~\bibnamefont {Minkov}},
  \bibinfo {author} {\bibfnamefont {Z.}~\bibnamefont {Yu}},\ and\ \bibinfo
  {author} {\bibfnamefont {H.}~\bibnamefont {Cao}},\ }\bibfield  {title}
  {\bibinfo {title} {Anderson localization of electromagnetic waves in three
  dimensions},\ }\href {https://doi.org/10.1038/s41567-023-02091-7} {\bibfield
  {journal} {\bibinfo  {journal} {Nature Physics}\ }\textbf {\bibinfo {volume}
  {19}},\ \bibinfo {pages} {1308} (\bibinfo {year} {2023})}\BibitemShut
  {NoStop}%
\bibitem [{\citenamefont {Ni}\ and\ \citenamefont {Volz}(2021)}]{Ni:2021aa}%
  \BibitemOpen
  \bibfield  {author} {\bibinfo {author} {\bibfnamefont {Y.}~\bibnamefont
  {Ni}}\ and\ \bibinfo {author} {\bibfnamefont {S.}~\bibnamefont {Volz}},\
  }\bibfield  {title} {\bibinfo {title} {Evidence of phonon anderson
  localization on the thermal properties of disordered atomic systems},\ }\href
  {https://doi.org/10.1063/5.0073129} {\bibfield  {journal} {\bibinfo
  {journal} {Journal of Applied Physics}\ }\textbf {\bibinfo {volume} {130}},\
  \bibinfo {pages} {190901} (\bibinfo {year} {2021})}\BibitemShut {NoStop}%
\bibitem [{\citenamefont {Orso}(2017)}]{Orso:2017aa}%
  \BibitemOpen
  \bibfield  {author} {\bibinfo {author} {\bibfnamefont {G.}~\bibnamefont
  {Orso}},\ }\bibfield  {title} {\bibinfo {title} {Anderson transition of cold
  atoms with synthetic spin-orbit coupling in two-dimensional speckle
  potentials},\ }\href {https://doi.org/10.1103/PhysRevLett.118.105301}
  {\bibfield  {journal} {\bibinfo  {journal} {Physical Review Letters}\
  }\textbf {\bibinfo {volume} {118}},\ \bibinfo {pages} {105301} (\bibinfo
  {year} {2017})}\BibitemShut {NoStop}%
\bibitem [{\citenamefont {Vollhardt}\ and\ \citenamefont
  {W\"olfle}(1980{\natexlab{a}})}]{Vollhardt:1980aa}%
  \BibitemOpen
  \bibfield  {author} {\bibinfo {author} {\bibfnamefont {D.}~\bibnamefont
  {Vollhardt}}\ and\ \bibinfo {author} {\bibfnamefont {P.}~\bibnamefont
  {W\"olfle}},\ }\bibfield  {title} {\bibinfo {title} {Anderson localization in
  $d\le 2$ dimensions: A self-consistent diagrammatic theory},\ }\href
  {https://doi.org/10.1103/PhysRevLett.45.842} {\bibfield  {journal} {\bibinfo
  {journal} {Phys. Rev. Lett.}\ }\textbf {\bibinfo {volume} {45}},\ \bibinfo
  {pages} {842} (\bibinfo {year} {1980}{\natexlab{a}})}\BibitemShut {NoStop}%
\bibitem [{\citenamefont {Vollhardt}\ and\ \citenamefont
  {W\"olfle}(1980{\natexlab{b}})}]{Vollhardt:1980ab}%
  \BibitemOpen
  \bibfield  {author} {\bibinfo {author} {\bibfnamefont {D.}~\bibnamefont
  {Vollhardt}}\ and\ \bibinfo {author} {\bibfnamefont {P.}~\bibnamefont
  {W\"olfle}},\ }\bibfield  {title} {\bibinfo {title} {Diagrammatic,
  self-consistent treatment of the {Anderson} localization problem in $d\le 2$
  dimensions},\ }\href {https://doi.org/10.1103/PhysRevB.22.4666} {\bibfield
  {journal} {\bibinfo  {journal} {Phys. Rev. B}\ }\textbf {\bibinfo {volume}
  {22}},\ \bibinfo {pages} {4666} (\bibinfo {year}
  {1980}{\natexlab{b}})}\BibitemShut {NoStop}%
\bibitem [{\citenamefont {Kroha}(1990)}]{Kroha:1990aa}%
  \BibitemOpen
  \bibfield  {author} {\bibinfo {author} {\bibfnamefont {J.}~\bibnamefont
  {Kroha}},\ }\bibfield  {title} {\bibinfo {title} {Diagrammatic
  self-consistent theory of anderson localization for the tight-binding
  model},\ }\href
  {https://doi.org/https://doi.org/10.1016/0378-4371(90)90055-W} {\bibfield
  {journal} {\bibinfo  {journal} {Physica A: Statistical Mechanics and its
  Applications}\ }\textbf {\bibinfo {volume} {167}},\ \bibinfo {pages} {231}
  (\bibinfo {year} {1990})}\BibitemShut {NoStop}%
\bibitem [{\citenamefont {Kroha}\ \emph {et~al.}(1990)\citenamefont {Kroha},
  \citenamefont {Kopp},\ and\ \citenamefont {W{\"o}lfle}}]{Kroha:1990ab}%
  \BibitemOpen
  \bibfield  {author} {\bibinfo {author} {\bibfnamefont {J.}~\bibnamefont
  {Kroha}}, \bibinfo {author} {\bibfnamefont {T.}~\bibnamefont {Kopp}},\ and\
  \bibinfo {author} {\bibfnamefont {P.}~\bibnamefont {W{\"o}lfle}},\ }\bibfield
   {title} {\bibinfo {title} {Self-consistent theory of anderson localization
  for the tight-binding model with site-diagonal disorder},\ }\href
  {https://doi.org/10.1103/PhysRevB.41.888} {\bibfield  {journal} {\bibinfo
  {journal} {Physical Review B}\ }\textbf {\bibinfo {volume} {41}},\ \bibinfo
  {pages} {888} (\bibinfo {year} {1990})}\BibitemShut {NoStop}%
\bibitem [{\citenamefont {Vollhardt}\ and\ \citenamefont
  {W\"olfle}(1992)}]{Vollhardt:1992aa}%
  \BibitemOpen
  \bibfield  {author} {\bibinfo {author} {\bibfnamefont {D.}~\bibnamefont
  {Vollhardt}}\ and\ \bibinfo {author} {\bibfnamefont {P.}~\bibnamefont
  {W\"olfle}},\ }\bibfield  {title} {\bibinfo {title} {Self-consistent theory
  of anderson localization},\ }in\ \href@noop {} {\emph {\bibinfo {booktitle}
  {Electronic Phase Transitions}}},\ \bibinfo {editor} {edited by\ \bibinfo
  {editor} {\bibfnamefont {W.}~\bibnamefont {Hanke}}\ and\ \bibinfo {editor}
  {\bibfnamefont {{\relax Yu}.~V.}\ \bibnamefont {Kopaev}}}\ (\bibinfo
  {publisher} {Elsevier Science Publishers B. V., Amsterodam},\ \bibinfo {year}
  {1992})\ Chap.~\bibinfo {chapter} {1}, pp.\ \bibinfo {pages}
  {1--78}\BibitemShut {NoStop}%
\bibitem [{\citenamefont {Velick{\'y}}\ \emph {et~al.}(1968)\citenamefont
  {Velick{\'y}}, \citenamefont {Kirkpatrick},\ and\ \citenamefont
  {Ehrenreich}}]{Velicky:1968aa}%
  \BibitemOpen
  \bibfield  {author} {\bibinfo {author} {\bibfnamefont {B.}~\bibnamefont
  {Velick{\'y}}}, \bibinfo {author} {\bibfnamefont {S.}~\bibnamefont
  {Kirkpatrick}},\ and\ \bibinfo {author} {\bibfnamefont {H.}~\bibnamefont
  {Ehrenreich}},\ }\bibfield  {title} {\bibinfo {title} {Single-site
  approximations in the electronic theory of simple binary alloys},\ }\href
  {https://doi.org/10.1103/PhysRev.175.747} {\bibfield  {journal} {\bibinfo
  {journal} {Physical Review}\ }\textbf {\bibinfo {volume} {175}},\ \bibinfo
  {pages} {747} (\bibinfo {year} {1968})}\BibitemShut {NoStop}%
\bibitem [{\citenamefont {Elliott}\ \emph {et~al.}(1974)\citenamefont
  {Elliott}, \citenamefont {Krumhansl},\ and\ \citenamefont
  {Leath}}]{Elliot:1974aa}%
  \BibitemOpen
  \bibfield  {author} {\bibinfo {author} {\bibfnamefont {R.~J.}\ \bibnamefont
  {Elliott}}, \bibinfo {author} {\bibfnamefont {J.~A.}\ \bibnamefont
  {Krumhansl}},\ and\ \bibinfo {author} {\bibfnamefont {P.~L.}\ \bibnamefont
  {Leath}},\ }\bibfield  {title} {\bibinfo {title} {The theory and properties
  of randomly disordered crystals and related physical systems},\ }\href
  {https://doi.org/10.1103/RevModPhys.46.465} {\bibfield  {journal} {\bibinfo
  {journal} {Rev. Mod. Phys.}\ }\textbf {\bibinfo {volume} {46}},\ \bibinfo
  {pages} {465} (\bibinfo {year} {1974})}\BibitemShut {NoStop}%
\bibitem [{\citenamefont {Vlaming}\ and\ \citenamefont
  {Vollhardt}(1992)}]{Vlaming:1992aa}%
  \BibitemOpen
  \bibfield  {author} {\bibinfo {author} {\bibfnamefont {R.}~\bibnamefont
  {Vlaming}}\ and\ \bibinfo {author} {\bibfnamefont {D.}~\bibnamefont
  {Vollhardt}},\ }\bibfield  {title} {\bibinfo {title} {Controlled mean-field
  theory for disordered electronic systems: Single-particle properties},\
  }\href {https://doi.org/10.1103/PhysRevB.45.4637} {\bibfield  {journal}
  {\bibinfo  {journal} {Physical Review B}\ }\textbf {\bibinfo {volume} {45}},\
  \bibinfo {pages} {4637} (\bibinfo {year} {1992})}\BibitemShut {NoStop}%
\bibitem [{\citenamefont {Jani{\v s}}\ and\ \citenamefont
  {Vollhardt}(1992)}]{Janis:1992ab}%
  \BibitemOpen
  \bibfield  {author} {\bibinfo {author} {\bibfnamefont {V.}~\bibnamefont
  {Jani{\v s}}}\ and\ \bibinfo {author} {\bibfnamefont {D.}~\bibnamefont
  {Vollhardt}},\ }\bibfield  {title} {\bibinfo {title} {Coupling of quantum
  degrees of freedom in strongly interacting disordered electron systems},\
  }\href {https://doi.org/10.1103/PhysRevB.46.15712} {\bibfield  {journal}
  {\bibinfo  {journal} {Physical Review B}\ }\textbf {\bibinfo {volume} {46}},\
  \bibinfo {pages} {15712} (\bibinfo {year} {1992})}\BibitemShut {NoStop}%
\bibitem [{\citenamefont {Velick{\'y}}(1969)}]{Velicky:1969aa}%
  \BibitemOpen
  \bibfield  {author} {\bibinfo {author} {\bibfnamefont {B.}~\bibnamefont
  {Velick{\'y}}},\ }\bibfield  {title} {\bibinfo {title} {Theory of electronic
  transport in disordered binary alloys: Coherent-potential approximation},\
  }\href {https://doi.org/10.1103/PhysRev.184.614} {\bibfield  {journal}
  {\bibinfo  {journal} {Physical Review}\ }\textbf {\bibinfo {volume} {184}},\
  \bibinfo {pages} {614} (\bibinfo {year} {1969})}\BibitemShut {NoStop}%
\bibitem [{\citenamefont {Khurana}(1990)}]{Khurana:1990aa}%
  \BibitemOpen
  \bibfield  {author} {\bibinfo {author} {\bibfnamefont {A.}~\bibnamefont
  {Khurana}},\ }\bibfield  {title} {\bibinfo {title} {Electrical conductivity
  in the infinite-dimensional hubbard model},\ }\href
  {https://doi.org/10.1103/PhysRevLett.64.1990} {\bibfield  {journal} {\bibinfo
   {journal} {Physical Review Letters}\ }\textbf {\bibinfo {volume} {64}},\
  \bibinfo {pages} {1990} (\bibinfo {year} {1990})}\BibitemShut {NoStop}%
\bibitem [{\citenamefont {Jani{\v s}}\ and\ \citenamefont
  {Vollhardt}(2001)}]{Janis:2001aa}%
  \BibitemOpen
  \bibfield  {author} {\bibinfo {author} {\bibfnamefont {V.}~\bibnamefont
  {Jani{\v s}}}\ and\ \bibinfo {author} {\bibfnamefont {D.}~\bibnamefont
  {Vollhardt}},\ }\bibfield  {title} {\bibinfo {title} {Conductivity of
  disordered electrons: Mean-field approximation containing vertex
  corrections},\ }\href {https://doi.org/10.1103/PhysRevB.63.125112} {\bibfield
   {journal} {\bibinfo  {journal} {Physical Review B}\ }\textbf {\bibinfo
  {volume} {63}},\ \bibinfo {pages} {125112} (\bibinfo {year}
  {2001})}\BibitemShut {NoStop}%
\bibitem [{\citenamefont {Jani{\v s}}\ \emph {et~al.}(2003)\citenamefont
  {Jani{\v s}}, \citenamefont {Koloren{\v c}},\ and\ \citenamefont {{\v S}pi{\v
  c}ka}}]{Janis:2003aa}%
  \BibitemOpen
  \bibfield  {author} {\bibinfo {author} {\bibfnamefont {V.}~\bibnamefont
  {Jani{\v s}}}, \bibinfo {author} {\bibfnamefont {J.}~\bibnamefont {Koloren{\v
  c}}},\ and\ \bibinfo {author} {\bibfnamefont {V.}~\bibnamefont {{\v S}pi{\v
  c}ka}},\ }\bibfield  {title} {\bibinfo {title} {Density and current response
  functions in strongly disordered electron systems: diffusion, electrical
  conductivity and einstein relation},\ }\href@noop {} {\bibfield  {journal}
  {\bibinfo  {journal} {European Physical Journal B}\ }\textbf {\bibinfo
  {volume} {35}},\ \bibinfo {pages} {77} (\bibinfo {year} {2003})}\BibitemShut
  {NoStop}%
\bibitem [{\citenamefont {Jani{\v s}}(2001)}]{Janis:2001ab}%
  \BibitemOpen
  \bibfield  {author} {\bibinfo {author} {\bibfnamefont {V.}~\bibnamefont
  {Jani{\v s}}},\ }\bibfield  {title} {\bibinfo {title} {Parquet approach to
  nonlocal vertex functions and electrical conductivity of disordered
  electrons},\ }\href {https://doi.org/10.1103/PhysRevB.64.115115} {\bibfield
  {journal} {\bibinfo  {journal} {Physical Review B}\ }\textbf {\bibinfo
  {volume} {64}},\ \bibinfo {pages} {115115} (\bibinfo {year}
  {2001})}\BibitemShut {NoStop}%
\bibitem [{\citenamefont {Jani{\v s}}\ and\ \citenamefont {Koloren{\v
  c}}(2005{\natexlab{a}})}]{Janis:2005aa}%
  \BibitemOpen
  \bibfield  {author} {\bibinfo {author} {\bibfnamefont {V.}~\bibnamefont
  {Jani{\v s}}}\ and\ \bibinfo {author} {\bibfnamefont {J.}~\bibnamefont
  {Koloren{\v c}}},\ }\bibfield  {title} {\bibinfo {title} {Mean-field theory
  of anderson localization: Asymptotic solution in high spatial dimensions},\
  }\href {https://doi.org/10.1103/PhysRevB.71.033103} {\bibfield  {journal}
  {\bibinfo  {journal} {Physical Review B}\ }\textbf {\bibinfo {volume} {71}},\
  \bibinfo {pages} {033103} (\bibinfo {year} {2005}{\natexlab{a}})}\BibitemShut
  {NoStop}%
\bibitem [{\citenamefont {Jani{\v s}}\ and\ \citenamefont {Koloren{\v
  c}}(2005{\natexlab{b}})}]{Janis:2005ab}%
  \BibitemOpen
  \bibfield  {author} {\bibinfo {author} {\bibfnamefont {V.}~\bibnamefont
  {Jani{\v s}}}\ and\ \bibinfo {author} {\bibfnamefont {J.}~\bibnamefont
  {Koloren{\v c}}},\ }\bibfield  {title} {\bibinfo {title} {Mean-field theories
  for disordered electrons: Diffusion pole and anderson localization},\ }\href
  {https://doi.org/10.1103/PhysRevB.71.245106} {\bibfield  {journal} {\bibinfo
  {journal} {Physical Review B}\ }\textbf {\bibinfo {volume} {71}},\ \bibinfo
  {pages} {245106} (\bibinfo {year} {2005}{\natexlab{b}})}\BibitemShut
  {NoStop}%
\bibitem [{\citenamefont {Jani{\v s}}\ and\ \citenamefont {Koloren{\v
  c}}(2004{\natexlab{a}})}]{Janis:2004aa}%
  \BibitemOpen
  \bibfield  {author} {\bibinfo {author} {\bibfnamefont {V.}~\bibnamefont
  {Jani{\v s}}}\ and\ \bibinfo {author} {\bibfnamefont {J.}~\bibnamefont
  {Koloren{\v c}}},\ }\bibfield  {title} {\bibinfo {title} {Conservation laws
  in disordered electron systems: Thermodynamic limit and configurational
  averaging},\ }\href@noop {} {\bibfield  {journal} {\bibinfo  {journal}
  {Physica Status Solidi (b)}\ }\textbf {\bibinfo {volume} {241}},\ \bibinfo
  {pages} {2032} (\bibinfo {year} {2004}{\natexlab{a}})}\BibitemShut {NoStop}%
\bibitem [{\citenamefont {Jani{\v s}}\ and\ \citenamefont {Koloren{\v
  c}}(2004{\natexlab{b}})}]{Janis:2004ab}%
  \BibitemOpen
  \bibfield  {author} {\bibinfo {author} {\bibfnamefont {V.}~\bibnamefont
  {Jani{\v s}}}\ and\ \bibinfo {author} {\bibfnamefont {J.}~\bibnamefont
  {Koloren{\v c}}},\ }\bibfield  {title} {\bibinfo {title} {Causality versus
  ward identity in disordered electron systems},\ }\href@noop {} {\bibfield
  {journal} {\bibinfo  {journal} {Modern Physics Letters B}\ }\textbf {\bibinfo
  {volume} {18}},\ \bibinfo {pages} {1051} (\bibinfo {year}
  {2004}{\natexlab{b}})}\BibitemShut {NoStop}%
\bibitem [{\citenamefont {Jani{\v s}}\ and\ \citenamefont {Koloren{\v
  c}}(2016)}]{Janis:2016aa}%
  \BibitemOpen
  \bibfield  {author} {\bibinfo {author} {\bibfnamefont {V.}~\bibnamefont
  {Jani{\v s}}}\ and\ \bibinfo {author} {\bibfnamefont {J.}~\bibnamefont
  {Koloren{\v c}}},\ }\bibfield  {title} {\bibinfo {title} {Conserving
  approximations for response functions of the fermi gas in a random
  potential},\ }\href@noop {} {\bibfield  {journal} {\bibinfo  {journal}
  {European Physical Journal B}\ }\textbf {\bibinfo {volume} {89}},\ \bibinfo
  {pages} {1434} (\bibinfo {year} {2016})}\BibitemShut {NoStop}%
\bibitem [{\citenamefont {Jani{\v s}}(1989)}]{Janis:1989aa}%
  \BibitemOpen
  \bibfield  {author} {\bibinfo {author} {\bibfnamefont {V.}~\bibnamefont
  {Jani{\v s}}},\ }\bibfield  {title} {\bibinfo {title} {Free-energy functional
  in the generalized coherent-potential approximation},\ }\href
  {https://doi.org/10.1103/PhysRevB.40.11331} {\bibfield  {journal} {\bibinfo
  {journal} {Physical Review B}\ }\textbf {\bibinfo {volume} {40}},\ \bibinfo
  {pages} {11331} (\bibinfo {year} {1989})}\BibitemShut {NoStop}%
\bibitem [{\citenamefont {Jani{\v s}}(2009)}]{Janis:2009aa}%
  \BibitemOpen
  \bibfield  {author} {\bibinfo {author} {\bibfnamefont {V.}~\bibnamefont
  {Jani{\v s}}},\ }\bibfield  {title} {\bibinfo {title} {Integrability of the
  diffusion pole in the diagrammatic description of noninteracting electrons in
  a random potential},\ }\href {https://doi.org/10.1088/0953-8984/21/48/485501}
  {\bibfield  {journal} {\bibinfo  {journal} {Journal of Physics: Condensed
  Matter}\ }\textbf {\bibinfo {volume} {21}},\ \bibinfo {pages} {485501}
  (\bibinfo {year} {2009})}\BibitemShut {NoStop}%
\bibitem [{\citenamefont {de~Almeida}\ and\ \citenamefont
  {Thouless}(1978)}]{deAlmeida:1978aa}%
  \BibitemOpen
  \bibfield  {author} {\bibinfo {author} {\bibfnamefont {J.~R.~L.}\
  \bibnamefont {de~Almeida}}\ and\ \bibinfo {author} {\bibfnamefont {D.~J.}\
  \bibnamefont {Thouless}},\ }\bibfield  {title} {\bibinfo {title} {Stability
  of the sherrington-kirkpatrick solution of a spin glass model},\ }\href@noop
  {} {\bibfield  {journal} {\bibinfo  {journal} {Journal of Physics A:
  Mathematical and General}\ }\textbf {\bibinfo {volume} {11}},\ \bibinfo
  {pages} {983} (\bibinfo {year} {1978})}\BibitemShut {NoStop}%
\end{thebibliography}


\end{document}